\documentclass{mystyle}
\copyrightyear{2011}
\pubyear{2011}

\usepackage{url}

\begin{document}
\firstpage{1}

\title[Machine Learning Pipeline for Discriminant Pathways Identification]{A MACHINE LEARNING PIPELINE FOR DISCRIMINANT PATHWAYS IDENTIFICATION}
\author[Barla \textit{et~al}]{
Annalisa Barla$^{(1)}$, 
Giuseppe Jurman$^{(2)}$, 
Roberto Visintainer$^{(2,3)}$
Margherita Squillario$^{(1)}$, 
Michele Filosi$^{(2,4)}$, 
Samantha Riccadonna$^{(2)}$ and
Cesare Furlanello$^{(2)}$}%\footnote{to whom correspondence should be addressed}}
\address{$^{1}$DISI, University of Genoa, via Dodecaneso 35, I-16146 Genova, Italy.\\
$^{2}$FBK, via Sommarive 18, I-38123 Povo (Trento), Italy.\\
$^{3}$DISI, University of Trento, via Sommarive 14, I-38123 Povo (Trento), Italy.\\
$^{4}$CIBIO, University of Trento, via Delle Regole 101, I-38123 Mattarello (Trento), Italy
}

\maketitle

\tableofcontents
\vskip.5cm
\hrule
\vskip.5cm

\begin{abstract}

\section{Motivation:}
Identifying the molecular pathways more prone to disruption during a pathological process is a key task in network medicine and, more in general, in systems biology.
\section{Results:}
In this work we propose a pipeline that couples a machine learning solution for molecular profiling with a recent  network comparison method. The pipeline can identify changes occurring between specific sub-modules of networks built in a case-control biomarker study, discriminating key groups of genes whose interactions are modified by an underlying condition. The proposal  is independent from the classification algorithm used. Three applications on genomewide data are presented regarding children susceptibility to air pollution and two neurodegenerative diseases: Parkinson's and Alzheimer's.
\section{Availability:}
Details about the software used for the experiments discussed in this paper are provided in the Appendix.

\section{Contact:} \href{furlan@fbk.eu}{furlan@fbk.eu}
\end{abstract}

\section{Introduction}
\label{sec:into}

Nowadays, it is widely accepted as a consolidated fact that most of the known diseases are of systemic nature: their phenotypes can be attributed to the breakdown of a rather complex set of molecular interaction among the cell's components rather than imputed to the misfunctioning of a single entity such as a gene.
A major aim of systems biology and, in particular, of its newly emerging discipline known as network medicine (\cite{barabasi11network}), is  the understanding of the cellular wiring diagram at all possible levels of organization (from transcriptomics to signalling) of the functional design, the molecular pathways being a typical example.
Such reconstruction is made feasible by the recent advances in the theory of complex networks (\textit{e.g.} \cite{strogatz01exploring,newman03structure,boccaletti06complex,newman10networks,buchanan10networks}) and, in particular, in the reconstruction algorithms for inferring networks topology and wiring starting from a collection of high-throughput measurements (\cite{he09reverse}). 
However, the tackled problem is hard (''a daunting task'', \cite{baralla09inferring}) and these methods are not flawless \cite{marbach10revealing}, due to many factors. Among them, underdeterminacy is a major issue (\cite{desmet10advantages}), and the ratio between the network dimension (number of nodes) and the number of available measurements to infer interactions plays a key role for the stability of the reconstructed structure. Although some effort has recently been put into facing this issue, the stability (and thus the reproducibility) of the process is still an open problem.

In this contribution we propose a pipeline for machine learning driven determination of the disruption of important molecular pathways induced or inducing a condition starting from microarray measurements in a case/control experimental design.
The problem of underdeterminacy in the inference procedure is avoided by focussing only on subnetworks, and the relevance of the studied pathways for the disease is judged in terms of discriminative relevance for the underlying classification problem. 
The profiling part of the pipeline, composed of a classifier and a feature selection method embedded within an adequate experimental procedure or Data Analysis Protocol (\cite{maqc10maqcII}), is used to rank the genes with the highest discriminative power. 
These genes undergo an enrichment phase (\cite{zhang05webgestalt,subramanian05gene}) to individuate the involved whole pathways to keep track of the established functional dependencies that would otherwise get lost by limiting the subnetwork analysis to the sole selected genes.
Finally, a network is inferred for both the case and the control samples on the selected pathways, and the two structure are compared to pinpoint the occurring differences and thus to detect the relevant pathway related variations.

A noteworthy point of this workflow is the independence from its ingredients: the classifier, the feature ranking algorithm, the enrichment procedure, the inference method and the network comparison function.
This last point is worth a comment: although already fruitfully used even in a biological context (\cite{sharan06modeling}), the problem of quantitatively comparing network (\textit{e.g.} using a metric instead of evaluating network properties) is a widely open issue affecting many scientific disciplines. 
As discussed in (\cite{jurman10introduction}), many classical distances (such as those of the edit family) have a relevant drawback in being local, that is focussing only on the portions of the network interested by the differences in the presence/absence of matching links. 
%On the other hand, more recent 
More recently, other metrics can overcome this problem so to consider the global structure of the compared topologies; among such distances, the spectral ones - based on the list of eigenvalues of the laplacian matrix of the underlying graph - are quite interesting, and, in particular, the Ipsen-Mikhailov (\cite{ipsen02evolutionary}) distance has been proven to be the most robust in a wide range of situations.

In what follows we will describe the newly introduced workflow in details, providing three 
examples of application in problems of biological interest: 
the first tasks concerns the transcriptomics consequences of exposition to environmental pollution on a cohort of children in Czech Republic,  the second one investigates
the molecular characteristics between Parkinson's disease (PD) at early and late stages and the
third regards the characterization of Alzheimer's disease (AD) at early and late stages.
To strenghten the support to our proposal, the two problems will be dealt with by using different experimental conditions, \textit{i.e.}, varying the employed algorithms throughout the various steps of the workflow. 
In both cases, biologically meaningful considerations can be drawn, consistent with previous findings, showing the effectiveness of the proposed procedure in the assessment of the occuring subnetwork variations.

%\begin{methods}
\section{System and Methods}
\label{sec:methods}

\begin{figure*}[!htb]
\centerline{\includegraphics[width=.9\textwidth]{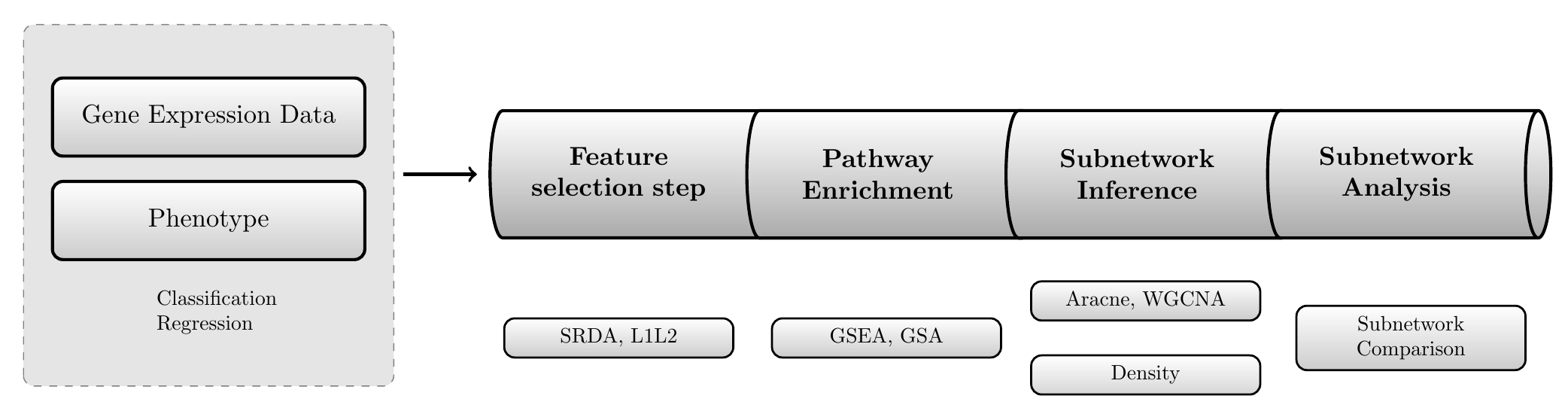}}
\caption{Schema of the analysis pipeline.}\label{fig:pipeline}
\end{figure*}

The proposed machine learning pipeline handles case/control transcription data through four main steps, from a profiling task output (a ranked list of genes) to the identification of discriminant pathways, see Figure~\ref{fig:pipeline}.
Alternative algorithms can be used at each step of the pipeline: as an example in the profiling part different  classifiers, regression or feature selection methods can be adopted.
In Section~\ref{sec:setup}  we describe the elementary steps used in the experiments. 

Formally, we are given a collection of $n$ subjects, each described by a $p$-dimensional vector $x$ of measurements.
Each sample is also associated with a phenotypical label $y=\{1,-1\}$, assigning it to a class (e.g. pollution vs. no-pollution in Section \ref{ssec:pollution}).
The dataset is therefore represented by an $n \times p$ gene expression data matrix $X$, where $p >> n$ and a corresponding labels vector $Y$.

The matrix $X$ is used to feed the profiling part of the pipeline. We choose a proper Data Analysis Protocol (\cite{maqc10maqcII}) for ensuring accurate and reproducible results and a prediction model. The model is built as a classifier (e.g., SRDA, \cite{cai08srda}) or a regression method (e.g., $\ell_1\ell_2$, \cite{DeMol:2009b}) coupled with a feature selection algorithm.
Thus, we obtain a ranked list of genes %$g_1,..., g_p$ 
from which we extract a gene signature $g_1, ..., g_k$ taking the top-$k$ most discriminant genes.
The choice is performed by finding a balance between the accuracy of the classifier and the stability of the signature (\cite{maqc10maqcII}).

Applying pathway enrichment techniques (\textit{e.g.}, GSEA or GSA, \cite{zhang05webgestalt,subramanian05gene}), we retrieve for each gene $g_i$ the corresponding whole pathway $p_i~=~\{h_1,...,h_t\}$, where the genes $h_j \not = g_i$ not necessarily belong to the original signature $g_1, ..., g_k$. Extending the analysis to all the $h_j$ genes of the pathway allows us to explore functional interactions that would otherwise get lost.

The subnetwork inference phase (\textit{e.g.}, WGCN or ARACNE, \cite{zhao10weighted, meyer08minet}) requires to reconstruct a network for each pathway $p_i$ by using the steady state expression data of the samples of each class $y$. The network inference procedure is limited to the sole genes belonging to the pathway $p_i$ in order to avoid the problem of intrinsic underdeterminacy of the task. As an additional caution against this problem, in the following experiments we limit the analysis to pathways having more than 4 nodes and less than 1000 nodes. For each $p_i$ and for each $y$, we obtain a real-valued adjacency matrix, which is then binarized by choosing a threshold on the correlation values.
This choice requires the construction of a binary adjacency matrix $N_{p_i,y,t_s}$ for each $p_i$, for each $y$ and for a grid of threshold values ${t_1, ..., t_T}$. For each value $t_s$ of the grid, we compute for each $p_i$ both the distance $D$ (e.g., the Ipsen-Mikhailov distance, see for details Section~\ref{ssec:distance}) between the case and control pathway graphs and the corresponding densities. We chose $t_s$ providing the best balance between the average distance across the pathways $p_i$ and the network density. For a fixed $t_s$ and for each $p_i$, we obtain a score $D(N_{p_i,1,t_s}, N_{p_i,-1,t_s})$ used to rank the pathways $p_i$.
As an additional scoring indicator for $g_1,...,g_k$, we also provide the difference between the weighted degree in the control ($y=-1$) and in the patient ($y=1$) network: $\Delta d(g_i)= d_{-1}(g_i)-d_{1}(g_i)$.
A final step of biological relevance assessment of the ranked pathways concludes the pipeline.

%\end{methods}

\section{Data Description}
\label{sec:data}

Section \ref{sec:discussion} describes three different experiments. 
In the first experiment we used
a genome-wide dataset created for investigating the effects of air pollution on children. 
In the second and third experiment we analyzed gene expression data on two neurodegenerative diseases: Parkinson's  (PD) and Alzheimer's (AD).
All the examples are based on publicly available datasets on the Gene Expression Omnibus (GEO). 

\subsection{Children susceptibility to air pollution}
%\label{ssec:data-pollution}
The first dataset (GSE7543) collects data of children living 
in two regions of the Czech Republic with different air pollution levels 
(\cite{vanLeeuwen08genomic,vanLeeuwen06genomewide}): 
23 children recruited in the polluted area of Teplice and 24 children living in the cleaner area of Prachatice. 
%The study was designed taking care of having similar populations from both regions according to age, gender and socioeconomic level. 
Blood samples were hybridized on Agilent Human 1A 22k oligonucleotide microarrays.  After normalization we retained  17564 features.
%We keep 17564 features, which is the subset of common probes for both cohorts of adult and child data.

\subsection{Clinical stages of Parkinson's disease} %\label{ssec:data-PD}
For PD we consider two publicly available datasets from GEO: GSE6613 (\cite{Scherzer:2007}) and GSE20295 (\cite{Zhang:2005}).
The former includes 22 controls and 50 whole blood samples from patients predominantly at early PD stages 
while the latter is composed of 53 controls and 40 patients with late stage PD.
Biological data were hybridized on Affymetrix HG-U133A platform, estimating the expression of 22215 probesets for each sample.
%Gene expressions were extracted from the .CEL files and normalized using the GC-RMA method  \cite{Wu:2004}. 

\subsection{Clinical stages of Alzheimer's disease}%\label{ssec:data-AD}
%AD early and late stages (GSE9770, GSE5281)
For AD we analyzed two GEO datasets: GSE9770 and GSE5281 (\cite{Liang:2010,Liang:2008}). 
The first includes 74 controls and 34 samples from non-demented patients with AD (since it is the earliest AD diagnosed, we will label it as early hereafter) and the second is composed of 74 controls and 80 samples from patients with late onset AD.
The samples were extracted  from six brain regions, differently susceptible to the disease: 
entorhinal cortex (EC), hippocampus (HIP), middle temporal gyrus (MTG), posterior cingulate cortex (PC), superior frontal gyrus (SFG) 
and primary visual cortex (VCX). 
The latter is known to be relatively spared by the disease, therefore we did not consider the samples within the VCX region. 
Overall, we analyzed 62 controls and 29 AD samples  for GSE9770 and 62 controls and 68 AD samples
for GSE5281.
Biological data were hybridized on Affymetrix HG-U133Plus2.0 platform, estimating the expression of 54713 probesets for each sample.

\section{Discussion}\label{sec:discussion}

\subsection{Air Pollution Experiment}\label{ssec:pollution}

The SRDA analysis of the air pollution dataset was performed within a $100\times5$-fold cross validation (CV) schema, producing a gene signature, characterizing the molecular differences between children in Teplice (polluted) and Prachatice (not polluted).
The  signature consists of 50 probesets, corresponding to 43 genes, achieving $76\%$ accuracy.%, corresponding to MCC=$0.74$. 

The enrichment analysis on the signature allowed a functional characterization of the relevant genes, identifying 11 enriched ontologies in GO (listed in Appendix Table~\ref{tab:poll1}). We then constructed the corresponding WGCN network for the 11 selected pathways for both cases and controls. 
Full details about the experiment are reported in Appendix Section~\ref{ssec:pollution-supp}. 

\begin{table}[ht]
\caption{Air Pollution Experiment: most important pathways ranked by the normalized Ipsen-Mikhailov distance $\hat{\epsilon}$. The Entrez gene symbol ID is also provided for the selected probesets $g_1,...,g_k$ in the corresponding pathway.\label{tab:Pollsummary}}
\centering{\begin{tabular}{ccl}
\toprule 
Pathway Code & $\hat{\epsilon}$ & \multicolumn{1}{c}{Gene Symbol} \\ 
\midrule
 GO:0043066 & 0.257 & \\ 
 GO:0001501 & 0.149 & MATN3 \\ 
 GO:0007399 & 0.093 & NRGN \\ 
 GO:0016787 & 0.078 & DHX32, CLC \\
 GO:0005516 & 0.076 & MYH1 \\ 
 GO:0007275 & 0.076 & FKHL18, HOXB8, OLIG1 \\ 
 GO:0006954 & 0.048 & PROK2\\ 
\botrule
\end{tabular}}{}
\end{table}

In Table \ref{tab:Pollsummary} we report the most biologically relevant pathways, ranked for decreasing normalized Ipsen-Mikhailov distance $\hat{\epsilon}$, which provides a measure of the structural distance between the networks inferred for the two classes. 
%\textit{i.e.} the element of the list are those mostly differentiated between cases and controls. 
%Finally, in Table \ref{tab:poll} (right) we also report the degree variation between cases and control classes for the genes in the signature.
The most disrupted pathway is  GO:0043066, \textit{i.e. apoptosis} followed by GO:0001501 \textit{i.e. skeletal development}.  
Since the children under study are undergoing natural development, especially physical changes of their skeleton, the high differentiation between cases and controls of the GO:0001501 and the involvement of pathway GO:0007275 \textit{i.e. developmental process} is biologically very sound.
Another relevant pathway is GO:0006954, representing the response to infection or injury caused by chemical or physical agents.
Several genes included in GO:0005516, ({\em i.e. calmodulin binding}) bind or interact with calmodulin, that is a calcium-binding protein involved in many essential processes, such as inflammation, apoptosis, nerve growth, and immune response. 
This is a key pathway that is linked with all the above mentioned terms as well as to GO:0007399, {\em i.e. nervous system development}, being one of the most stimulated pathways together with GO:0001501. 

%Table  \ref{tab:Pollsummary} also shows the genes that most 
%sensibly change their connection degree, that is, the strenght of their interactions within the pathway.

As described in Section \ref{sec:methods} the pipeline also provides a score $\Delta d$ of the variation of the number of interactions for  $g_1,...,g_k$. The full list is provided in Appendix Table~\ref{tab:poll2}, here we discuss a subset of the most biologically relevant genes.

FKHL18, HOXB8, PROK2, DHX32, MATN3 are directly involved in the development.
CLC is a  key element in the inflammation and immune system. 
OLIG1 is a transcription factor that works in the oligodendrocytes within the brain. NRGN binds calcium and is a target for thyroid hormones in the brain.
Finally,  MYH1 encodes for myosin that is a major contractile protein that forms striated, smooth and non-muscle cells. MYH1 isoforms show expression that is spatially and temporally regulated during development.

\begin{figure}[!h]
\begin{center}
\begin{tabular}{cc}
%\centerline{
\includegraphics[width=0.15\textwidth]{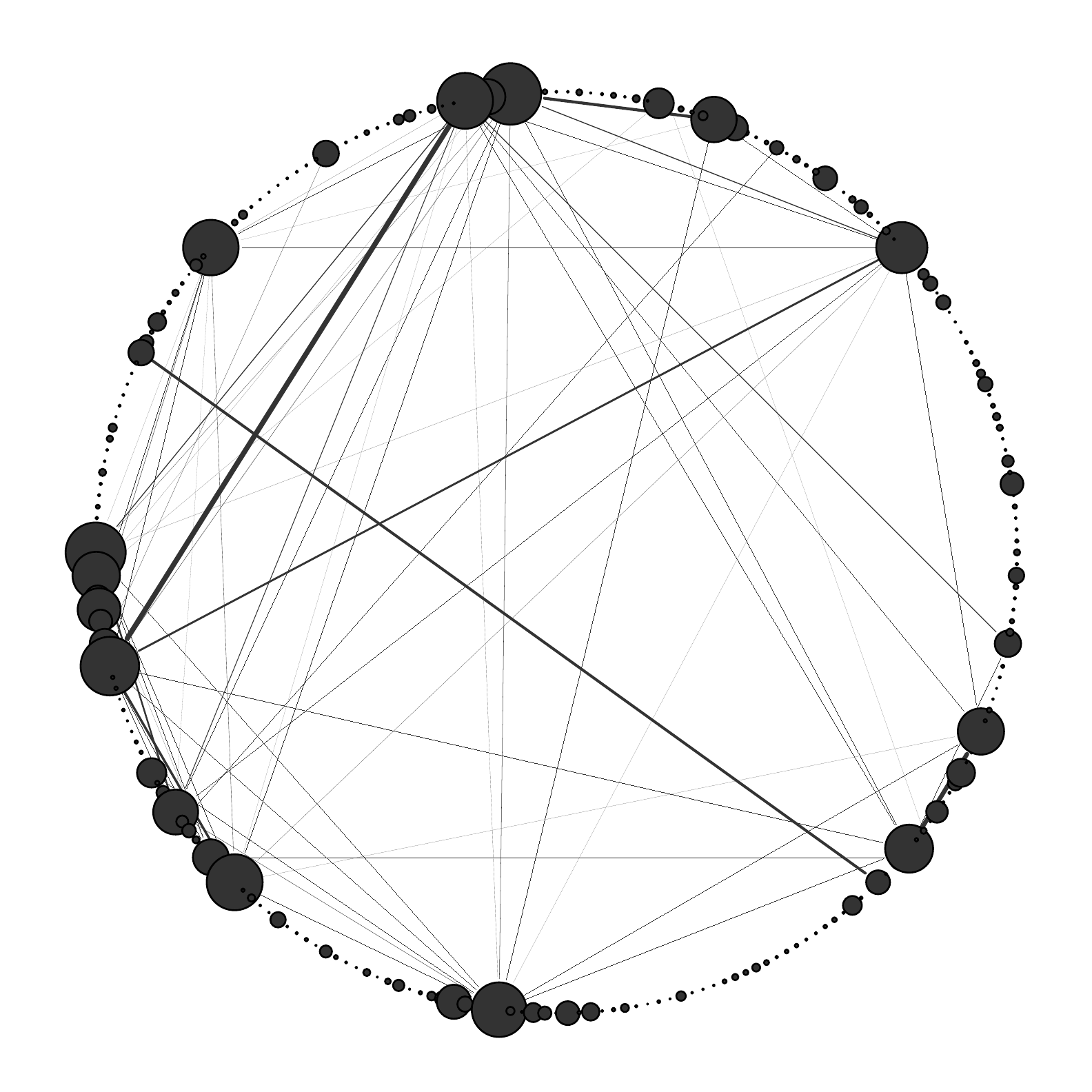} &
\includegraphics[width=0.15\textwidth]{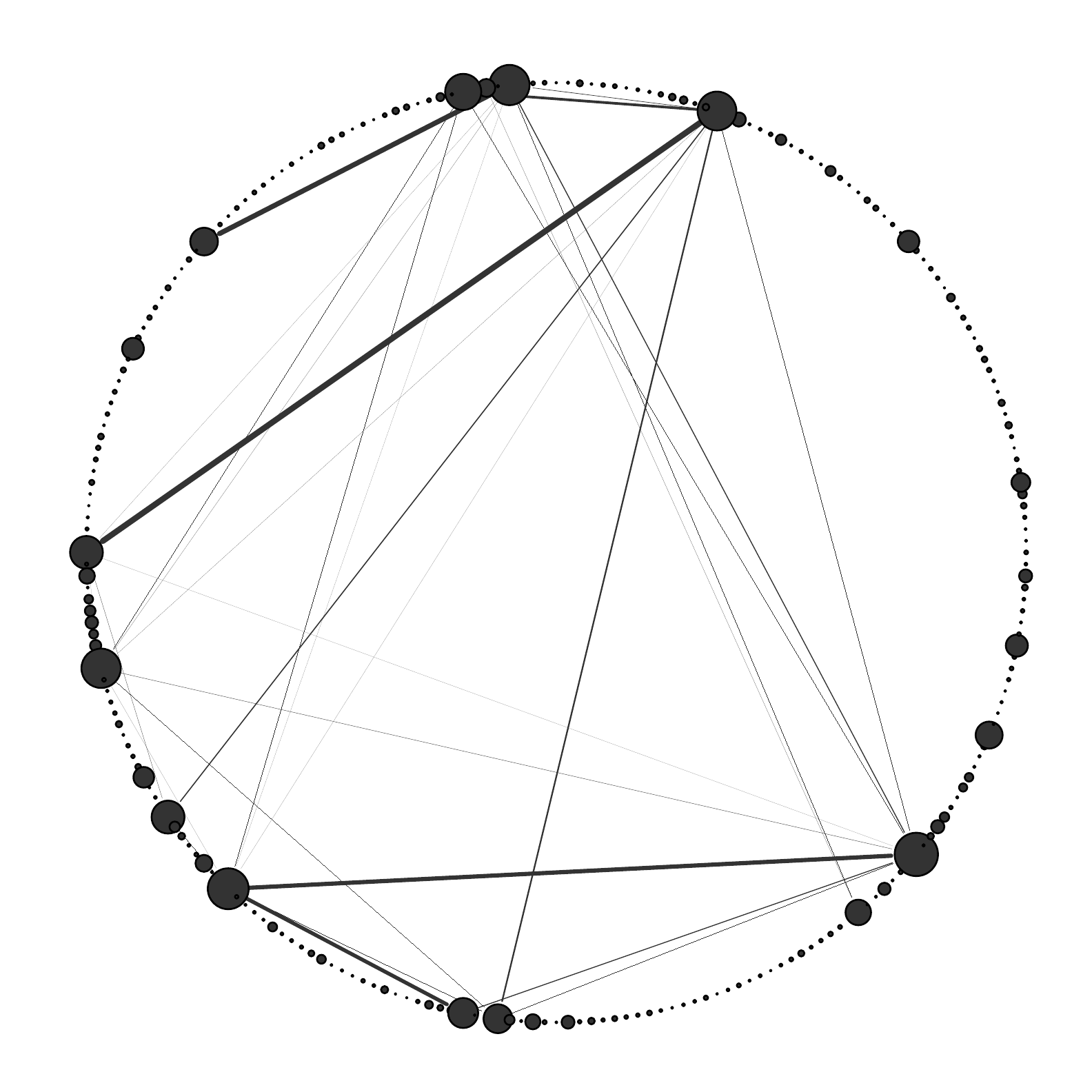} \\
%}
(a) Prachatice & (b)  Teplice\\
\end{tabular}
\caption{Networks of the pathway GO:0007399 (\textit{nervous system development}) for Prachatice children (a) compared with  Teplice children (b). 
Node diameter is proportional to the degree, and edge width is proportional to connection strength (estimated correlation).}\label{fig:AirNetworks}
\end{center}
\end{figure}

Figure \ref{fig:AirNetworks} shows the network of the  GO:0007399 pathway,  related to the nervous system development in the two cohorts. It is clear that several connections among the genes within this pathway  are missing  in the subjects living in the polluted area (Teplice). Therefore the nervous system development in these children is potentially at risk compared to those living in the not polluted city (Prachatice).

\subsection{Parkinson's Disease Experiment}
\label{ssec:PD}
The $\ell_1\ell_2$ analysis of the two PD datasets was performed respectively within a $9$-fold nested CV loop for the early PD and $8$-fold nested CV for the late PD. The early PD signature consists of 77 probesets, mapped on 70 genes, and associated to $62\%$ accuracy. The late stage signature is composed of 94 probesets corresponding to 90 genes and achieving $80\%$ accuracy. 

\begin{table}[ht]
\caption{PD: most important pathways ranked by normalized Ipsen-Mikhailov distance $\hat{\epsilon}$. The Entrez gene symbol ID is also provided for the selected probesets $g_1,...,g_k$ in the corresponding pathway. In bold, common elements between early and late stage PD.\label{tab:PDsummary}}
\centering{\begin{tabular}{lccl}
\toprule 
&Pathway Code & $\hat{\epsilon}$ & \multicolumn{1}{c}{Gene Symbol} \\ 
\midrule
PD early & GO:0005506 & 0.38 & HBB \\
& GO:0006952 & 0.37 & DEFA1/DEFA3 \\
& GO:0045087 & 0.36 & \\
& GO:0042802 & 0.33 & \textbf{SYT1} \\
& GO:0042802 & 0.33 & \textbf{HSPB1} \\
& GO:0006955 & 0.31 & DEFA1/DEFA3 \\
& GO:0006950 & 0.28 & \textbf{HSPB1} \\
& GO:0020037 & 0.26 & HBB \\
& GO:0005938 & 0.26 & \textbf{MYH10} \\
& GO:0005856 & 0.24 & \textbf{VCL}\\
& GO:0005856 & 0.24 & \textbf{HSPB1}\\
& \textbf{GO:0003779} & 0.23 & \textbf{MYH10}, \textbf{VCL}\\
& GO:0030097 & 0.15 & \\
& GO:0009615 & 0.14 & \\
& GO:0009615 & 0.14 & DEFA1/DEFA3\\
& GO:0051707 & 0.00 & \\
\midrule
PD late & GO:0019226 & 0.31 & \\
& GO:0007611 & 0.16 & \\ 
& GO:0042493 & 0.15 & \\
& GO:0009725 & 0.11 & \\
& GO:0030424 & 0.10 & \textbf{MYH10} \\
& GO:0007267 & 0.09 & TAC1 \\
& GO:0005516 & 0.09 & \textbf{MYH10}, \textbf{SYT1}, RGS4\\
& GO:0005096 & 0.09 & RGS4 \\
& GO:0007610 & 0.08 & \\
& \textbf{GO:0003779} & 0.08 & \textbf{MYH10}, \textbf{VCL}\\
& GO:0005624 & 0.08 & SLC18A2\\
& GO:0045202 & 0.08 & \textbf{SYT1}\\
& GO:0003924 & 0.07 & CDC42 \\
& GO:0006928 & 0.07 & HSBP1, \textbf{VCL}\\
& GO:0042995 & 0.07 & CDC42\\
& GO:0007268 & 0.06 & TAC1, \textbf{SYT1} \\
& GO:0043234 & 0.06 & \textbf{VCL}\\
& GO:0005525 & 0.05 & CDC42\\
& GO:0006412 & 0.05 & RPS4Y\\
& GO:0006836 & 0.05 & SLC18A2, SLC6A3 \\
& GO:0043005 & 0.05 & \textbf{MYH10}, \textbf{SYT1} \\
& GO:0043025 & 0.04 & \textbf{MYH10} \\
& GO:0042221 & 0.00 & \\
& GO:0009266 & 0.00 &  \\
& GO:0014070 & 0.00 & \\
\botrule
\end{tabular}}{}
\end{table}

%
%The intersection between the signatures resulted in four common genes: 
%XIST, RPS4Y1, DEFA1/DEFA3, HLA DQB1. %and HBB, PMS2L1/PMS2L2, SCAMP1, XIST for AD. 
%In particular, XIST  was recently  found, together with RPS4Y1, to be expressed in a subset of neurons as part
%of a group of gender-speciÞc genes differentially expressed in dorsolateral prefrontal cortex, anterior cingulate cortex and cerebellum. 
%For each signature, the  enrichment analysis identified relevant enriched nodes 
%either specific or common between early and late PD. 
%{\bf The common pathways have  a very general meaning (e.g. 
%\textit{intracellular, cytoplasm, negative regulation of  biological process}). 
%The specific ones for the early stage PD concern the immune system, the response to 
%stimulus (i.e. \textit{stress, chemicals or other organism like virus}), the regulation of 
%metabolic processes, the biological quality and cell death. 
%The specific pathways for late stage are related to the nervous system (e.g. \textit{neurotransmitter transport, 
%transmission of nerve impulse, learning or memory}) and 
%to  response to stimuli (e.g. \textit{behavior, temperature, organic substances, drugs or endogenous stimuli}). }

Applying ARACNE, we constructed the relevance network for both cases and controls for the 35 enriched pathways for late stage PD case and 42 pathways for early stage PD. Table \ref{tab:PDsummary} reports the most biologically relevant pathways, ranked for decreasing normalized Ipsen-Mikhailov distance $\hat{\epsilon}$. The full list of the analyzed pathways is provided as Appendix Table~\ref{tab:park}.

Having characterized the functional alteration of pathways for both early and late stage PD, we attempt a comparative
analysis of the outcome, commenting the most meaningful results from the biological viewpoint.
We expected some common pathways between the two stages, especially within pathways that represent
general processes and functions, but as commented in Section \ref{sec:methods} the pipeline does not consider 
pathways having more that 1000 nodes, hence  discarding the general terms in the GO.
Indeed, the only common pathway is GO:0003779, \textit{i.e. actin binding}. 
Actin  participates in many important cellular processes, including muscle contraction, cell motility, cell division and cytokinesis, 
vescicle and organelle movement, cell signaling. % and the establishment and maintenance of cell junctions and shapes.
Clearly, this term is strictly associated to the most evident movement-related symptoms in PD, including
shaking, rigidity, slowness of movement and difficulty with walking and gait.

In both early and late PD we note some alteration within the biological process class of \textit{response to stimulus}.
In the early PD list we identified GO:0006950  \textit{i.e. response to stress}, GO:0009615 \textit{i.e. response to virus} and GO:0051707 
 \textit{i.e. response to other organism}. In the late PD list we found GO:0042493, \textit{i.e. response to drug}, 
GO:0009725 \textit{i.e. response to hormone stimulus}, GO:0042221 \textit{i.e. response to chemical stimulus}, GO:0014070 {\em
i.e. response to organic cyclic substance} and GO:0009266 \textit{response to temperature stimulus}. 

\begin{figure}[!h]
\begin{center}
\begin{tabular}{cc}
%\centerline{
\includegraphics[width=0.15\textwidth]{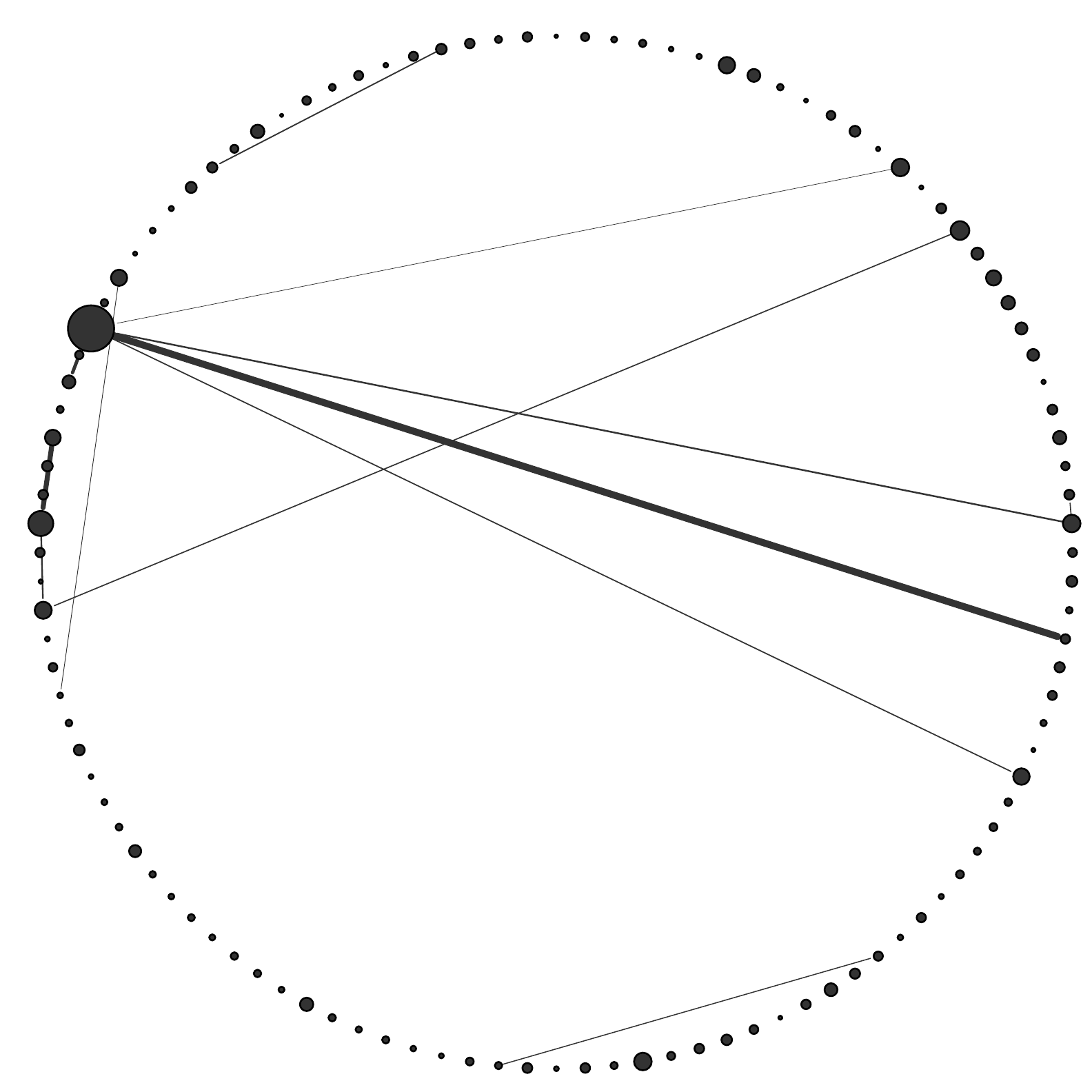} &
\includegraphics[width=0.15\textwidth]{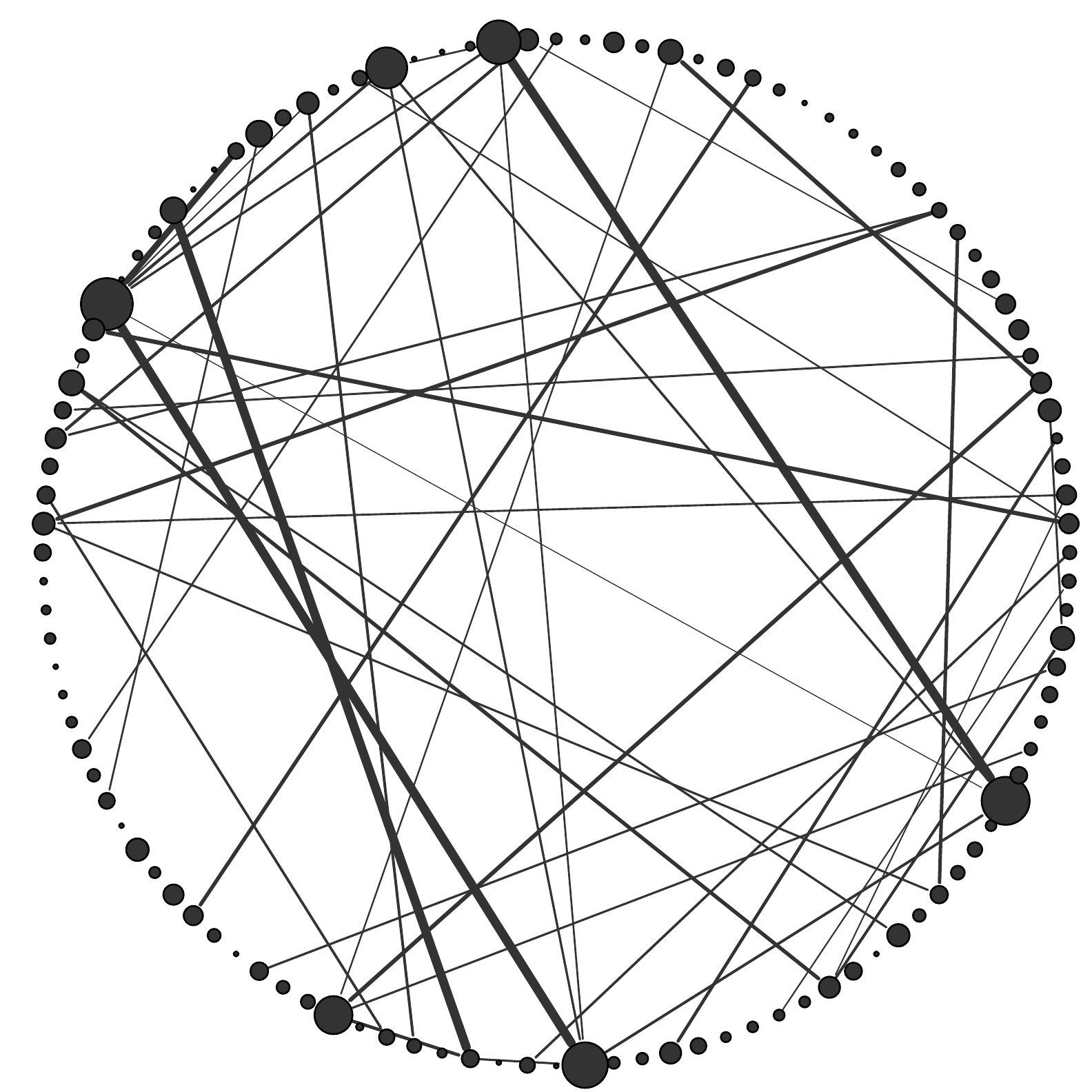} \\
%}
(a) early PD patients & (b) controls\\
\end{tabular}
\caption{Networks of the pathway GO:0045087 (\textit{innate immune response}) for PD early development patients (a) compared with healty subjects (b). Node diameter is proportional to the degree, and 
edge width is proportional to connection strength (estimated correlation).}\label{fig:PDnetwork}
\end{center}
\end{figure}

The pathways specific to  early PD show a great involvement of the immune system, which is greatly stimulated by inflammation especially located in particular brain regions (mainly \textit{substantia nigra}). 
Indeed, we identified: 
GO:0006952 \textit{i.e. defense response}, 
GO:0045087 \textit{i.e. innate immuno response} also visualized in Figure~\ref{fig:PDnetwork}, GO:0006955 \textit{i.e. immune response} 
and GO:0030097 \textit{i.e. hemopoiesis}.

From Figure \ref{fig:PDnetwork} it is clear that since the early stages of PD  the innate immune system is severly compromised: the body  is highly subjected to the invasion and proliferation of microbes (like bacteria or viruses), resulting in a debilitated organism,   less effective in fighting the consequent inflammation.

In late stage PD, we detected several differentiated terms related to the Central Nervous System. Among others, we mention: GO:0019226 \textit{i.e. transmission of nerve impulse}, GO:0007611 \textit{i.e. learning or memory}, GO:0007610 \textit{i.e.  behavior} and GO:007268 \textit{i.e. synaptic transmission}. 
These findings are fitting the late stage PD scenario, where cognitive and behavioral problems may arise with dementia.

%%%geni

%Common genes
Table \ref{tab:PDsummary}, and Appendix Table \ref{tab:park2early} and \ref{tab:park2late} report the genes belonging to the most relevant pathways for early and late PD, respectively. 
%Some of the genes are repeated one or more time within each table because they participate in several pathways. 

Four common genes were identified between early and late stage PD: MYH10, SYT1, VCL and  HSPB1.
MYH10 is involved in several pathways: {\em acting binding} and {\em calmodulin binding}, {\em neural cell body}, {\em neuron projection} and
{\em cell cortex}. 
These pathways indicate that the damage mostly occurs in the neurons and especially the actin binding and the cell cortex affect the cytoskeleton and the muscular tissue. At the same time, the calmodulin binding pathway indicates that other preprocesses, related to calmodulin and relevant for PD, might be damaged. These processes are related to the inflammation, metabolism, apoptosis, smooth muscle contraction, intracellular movement, short-term, long-term memory, nerve growth and the immune response. Moreover, it is known that MYH10 is involved in the regulation of the actin cytoskeleton pathways but also in that ones related to the axon guidance. Mutations in this gene are known to be present in disease phenotypes affecting the heart and the brain (\cite{Kim:2005p17378}). 
The synaptotagmin SYT1, also involved in the calmoduling binding, is an integral membrane protein of synaptic vesicles thought to serve as Ca(2+) sensor in the process of vesicular trafficking and exocytosis. Calcium binding to SYT1 participates in triggering neurotransmitter release at the synapse. This protein is therefore involved in the synaptic transmission and it predominantly  works in the neuron projections and synapses. 
Vinculin (VCL) is a cytoskeletal protein associated with cell-cell and cell-matrix junctions, where it is thought to function as one of several interacting proteins involved in anchoring F-actin to the membrane. Defects in VCL are the cause of cardiomyopathy dilated type 1W. This protein is involved in cell motility, proliferation and differentiation but also in smooth muscle contraction, inflammation and immune surveillance. VCL is located on a locus of chromosome 10 strongly associated with late onset AD (\cite{Grupe:2006p17379}).
HSPB1 is a heat shock protein induced by environmental stress and developmental changes. The encoded protein is involved in stress resistance and actin organization and translocates from the cytoplasm to the nucleus upon stress induction. This translocation occurs in order to modulate SP1-dependent transcriptional activity to promote neuronal protection  (\cite{Friedman:2009p17380}).
Furthermore, defects in this gene cause  two neurophatic diseases (i.e. Charcot-Marie-Tooth disease type 2F and distal hereditary motor neuropathy).

%Parkinson Early top genes
Beside the common genes, early stage PD is characterized by several meaningful genes. 
HBB encodes for hemoglobin beta that, together with another hemoglobin beta and two hemoglobin alpha, forms the adult hemoglobin. The work of \cite{Atamna:2006p17383} shows that the binding of Abeta to the heme group (hemoglobins bond to iron) supports a unifying mechanism by which excessive Amyloid-beta (Abeta) induces heme deficiency, causes oxidative damage to macromolecules, and depletes specific neurotransmitters. Althought Abeta is a known marker for AD, 
a recent publication also places it within a panel of PD biomarkers (\cite{Shi:2011p17384}).
DEFA1 and DEFA3 are both defensins, a family of microbicidal and cytotoxic peptides thought to be involved in host defense. They are abundant in the granules of neutrophils and also found in the epithelia of mucosal surfaces such as those of the intestine, respiratory tract, urinary tract, and vagina. Recently, \cite{Andrianov:2007p17385} presented some evidence for the recruitment of defensins in communication between the immune and nervous systems in the frog.

%Parkinson Late top genes
Among the genes specific to the late PD, we note CDC42, a GTPase of the Rho-subfamily, which regulates signaling pathways controlling diverse cellular functions including cell morphology, migration, endocytosis and cell cycle progression. 
Through the interaction with other proteins, CDC42 is known to regulate actin polymerization constituent both of the cytoskeleton and of the muscle cells. 
SLC6A3 is a dopamine transporter which is a member of the sodium- and chloride-dependent neurotransmitter transporter family. This gene is associated with Parkinsonism-dystonia infantile (\cite{Kurian:2009p17382}).
Other significant genes within Table \ref{tab:PDsummary} are: TAC1, RGS4, SLC18A2 and RPS4Y.

\subsection{Alzheimer's Disease Experiment}
\label{ssec:AD}

Classification and feature selection via $\ell_1\ell_2$, performed within a 9-fold nested CV schema for AD early and 8-fold for AD late, gives respectively 90\% accuracy and 95\% with 50 probesets for both cases.

\begin{figure}[!h]
\begin{center}
\begin{tabular}{cc}
%\centerline{
\includegraphics[width=0.15\textwidth]{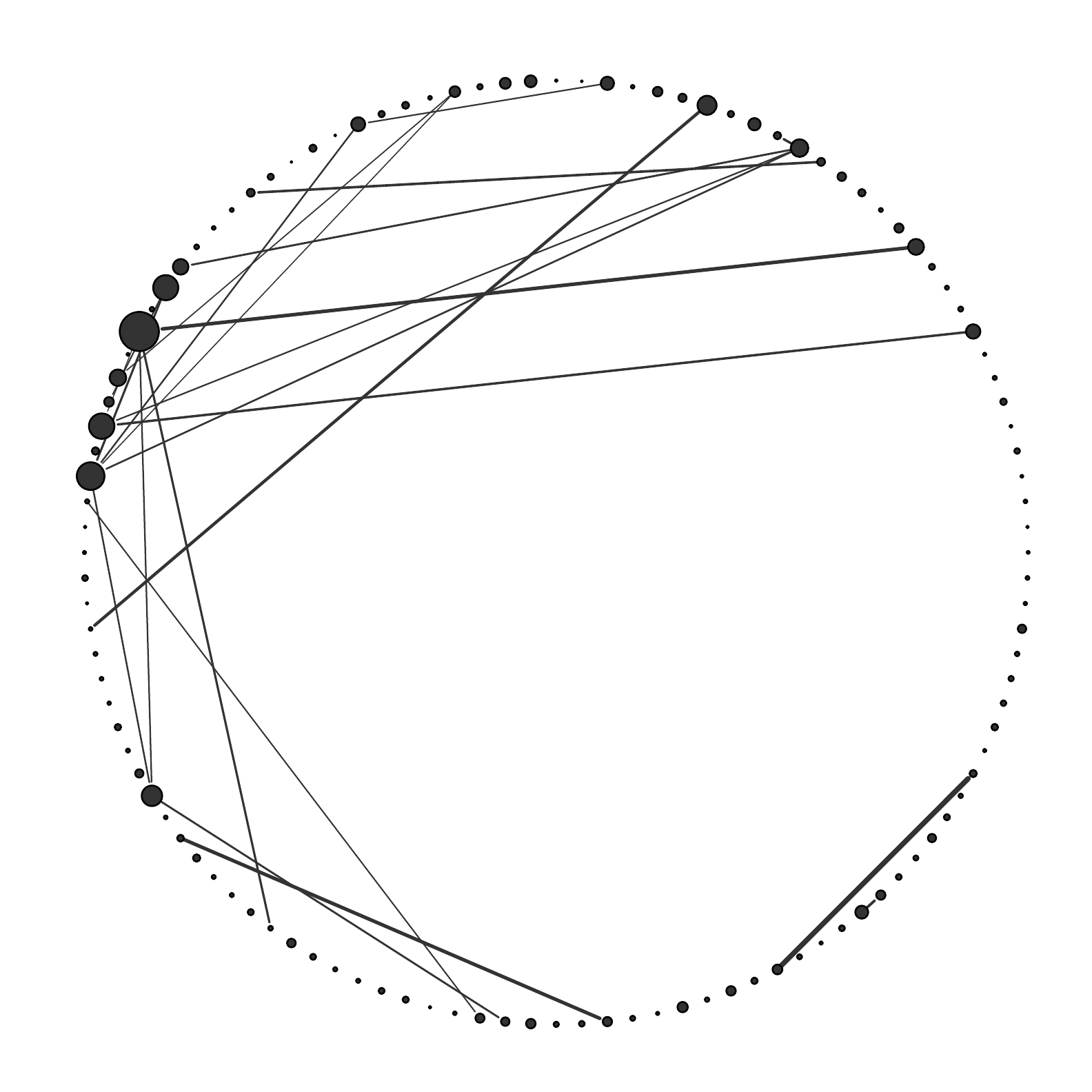} &
\includegraphics[width=0.15\textwidth]{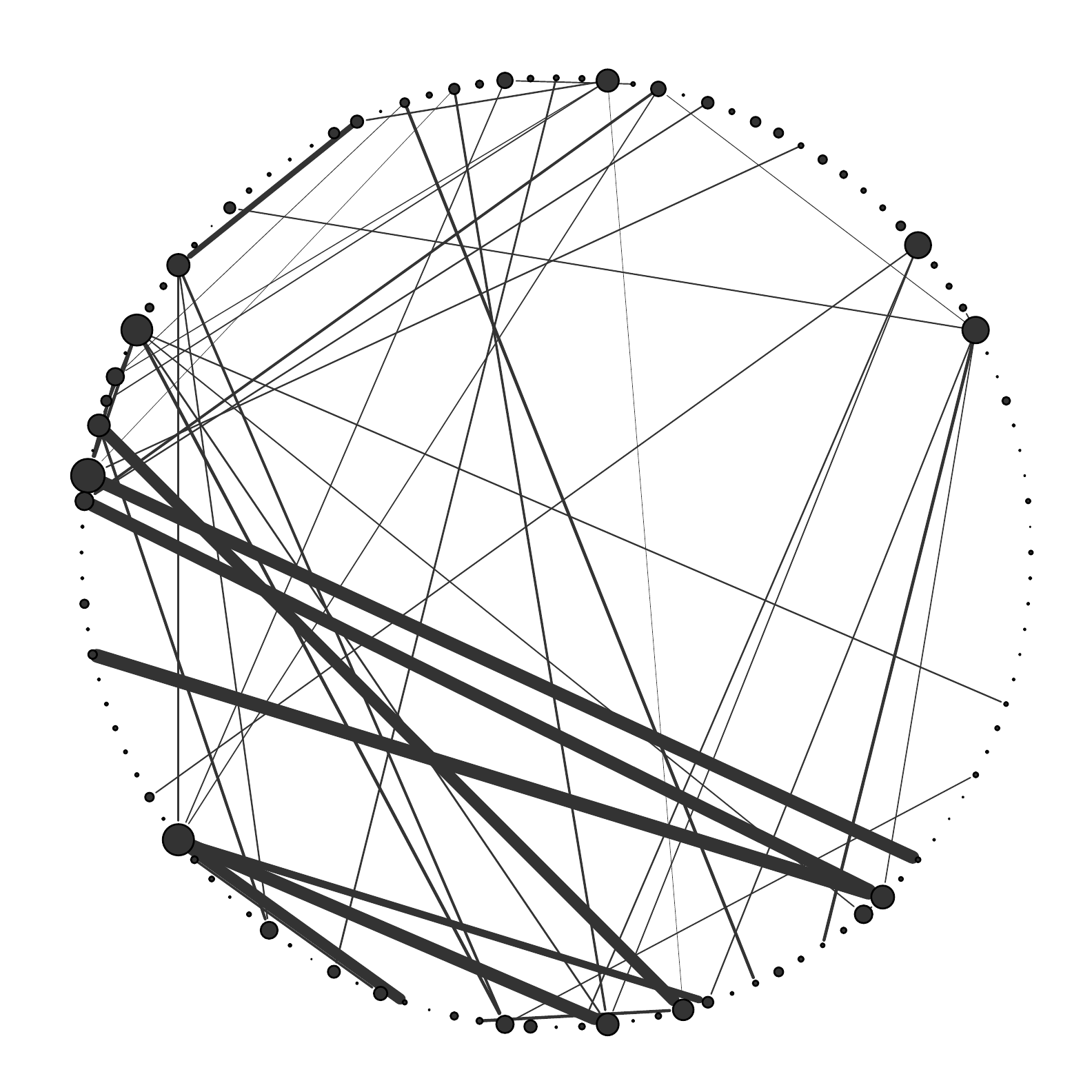} \\
%}
(a) early AD patients & (b) controls\\
\end{tabular}
\caption{Networks of the pathway GO:0019787 for AD early development patients (a) compared with healthy subjects (b). Node diameter is proportional to the degree, and 
edge width is proportional to connection strength (estimated correlation).}\label{fig:ADnetwork}
\end{center}
\end{figure}

We apply in the AD case the same network analysis strategy as in the PD experiment inferring for both cases and controls 51 selected pathways for early stage AD and 34 for late stage AD. The full list of reconstructed pathways is reported in Table~\ref{tab:alz}. In Table \ref{tab:ADsummary} we summarize the main findings discussed hereafter.

Similarly to the PD analysis, we attempt a comparative analysis of the outcome for early and late stage AD having characterized the functional alteration of pathways for the two AD stages and comment the most meaningful results from the biological viewpoint.

\begin{table}[ht]
\caption{AD: most important pathways ranked by normalized Ipsen-Mikhailov distance $\hat{\epsilon}$. The Entrez gene symbol ID is also provided for the selected probesets $g_1,...,g_k$ in the corresponding pathway. In bold, common pathways between early and late stage AD.\label{tab:ADsummary}}
\centering{\begin{tabular}{lccl}
\toprule 
&Pathway Code & $\hat{\epsilon}$ & \multicolumn{1}{c}{Gene Symbol} \\ 
\midrule
AD early & \textbf{GO:0042598} & 0.21 & \\
& GO:0019787 & 0.16 & UBE2D3 \\
& GO:0007417 & 0.10 & MPB \\
& GO:0001508 & 0.14 &\\
& GO:0051246 & 0.15 & UBE2D3 \\
& GO:0016874 & 0.12 & UBE2D3 \\
& GO:0004842 & 0.11 & UBE2D3 \\
& GO:0005768 & 0.08 & EGFR \\
& GO:0016567 & 0.07 & UBE2D3 \\
& GO:0050877 & 0.06 & \\
& GO:0042552 & 0.05 & \\
& \textbf{GO:0008015} & 0.04 & \\
& GO:0042391 & 0.04 & \\
& GO:0007399 & 0.04 & NTRK2 \\
& GO:0046982 & 0.03 & EGFR \\
& GO:0006633 & 0.02 & PTGDS\\
& \textbf{GO:0019226} & 0.00 & Ê\\
& \textbf{GO:0000267} & 0.00 & \\
\midrule
AD late &GO:0040012 & 0.36 & SNCA \\
& \textbf{GO:0042598} & 0.23 & \\
& \textbf{GO:0019226} & 0.12 & Ê\\ 
&GO:0030334 & 0.10 & \\
&GO:0045892 & 0.09 & SPEN\\
&GO:0042493 & 0.06 & SNCA \\ 
&GO:0042127 & 0.05 & \\
&GO:0008283 & 0.04 & CAT \\
&GO:0005215 & 0.03 & XK \\
&GO:0008217 & 0.03 & HBD\\
&GO:0007601 & 0.03 & \\
&GO:0007268 & 0.03 & \\
&GO:0007610 & 0.03 & \\
&GO:0008289 & 0.03 & \\
& \textbf{GO:0008015} & 0.02 & \\ 
&GO:0016564 & 0.02 & SPEN, ATXN1 \\
&GO:0008284 & 0.02 & \\
&GO:0008285 & 0.02 & EIF2AK1 \\
&GO:0020037 & 0.02 & EIF2AK1, CAT, HBD \\
& \textbf{GO:0000267} & 0.00 & \\
&GO:0050890 & 0.00 & \\
\botrule
\end{tabular}}{}
\end{table}

Four common pathways were identified: GO:0019226 \textit{i.e. transmission of nerve impulse}, GO:0008015 \textit{i.e. blood circulation}, 
GO:0000267 \textit{i.e. cell fraction} and GO:0042598 \textit{i.e. vesicular fraction}.

%%nodi AD early
The majority of pathways characterizing early stages of AD are related to the nervous system, and  the blood.
Among the nervous system related pathways the most damaged are: GO:0007399 \textit{i.e. nervous system development}, GO:0007417 \textit{i.e. central nervous system development}, 
GO:0042391 \textit{i.e. regulation of membrane potential}, GO:0042552 \textit{i.e. myelination}, GO:0050877 \textit{i.e. neurological system process},  GO:0001508 \textit{i.e. regulation of action potential} and GO:0019226 \textit{i.e. transmission of nerve impulse}.

%%nodi AD late
The majority of the pathways characterizing late stage AD are related to the cell, to the nervous system and to the response of the organism to various stimuli, see Table \ref{tab:ADsummary} and \ref{tab:alz}. Among the pathways centered on the cell, mentioned in descending order based on the numerosity of the genes, there are: GO:0008283 \textit{i.e. cell proliferation}, GO:0008283 \textit{i.e. negative regulation of cell proliferation}, GO:0008284 \textit{i.e. positive regulation of cell proliferation}, GO:0042127 \textit{i.e. regulation of cell proliferation}, GO:0030334 \textit{i.e. regulation of cell migration}. The pathways related to the nervous system  are: GO:0007268 \textit{i.e. synaptic transmission}, GO:0007610 \textit{i.e. behavior}, GO:0050890 \textit{i.e. cognition}. Other relevant nodes are those related to the transcription regulation (GO:0016564, GO:0045892), the visual perception (GO:0007601), and the heme and lipid binding (i.e. GO:0020037, GO:0008289).

%% geni AD early 
The genes characterizing the early stage AD are reported in Table \ref{tab:ADsummary} and \ref{tab:alz2early}. 
%Among these the most frequently repeated genes are HBB and UBE2D3.
%The relevance of HBB both in AD and PD have been already explained in the Parkinson's paragraph.
UBE2D3 is an ubiquitin,  targeting abnormal or short-lived proteins for degradation. 
It is a member of the E2 ubiquitin-conjugating enzyme family. This enzyme functions in the ubiquitination of the tumor-suppressor protein p53.
It is also involved in several signaling pathways  (BMP, TGF-$\beta$, TNF-$\alpha$/NF-kB and in the immune system), in the protein processing in the endoplasmatic reticulum.
PTGDS is an enzyme that catalyzes the conversion of prostaglandin H2 (PGH2) to postaglandin D2 (PGD2). It functions as a neuromodulator as well as a trophic factor in the central nervous system and it is also involved in smooth muscle contraction/relaxation and is a potent inhibitor of platelet aggregation. This gene is preferentially expressed in brain. Quantifying the protein complex of PGD2 and TTR in CSF may be useful in the diagnosis of AD, possibly in the early stages of the disease (\cite{Lovell:2008p17386}).
EGFR is a transmembrane glycoprotein that is a member of the protein kinase superfamily. This protein is a receptor for members of the epidermal growth factor family that binds to epidermal growth factor. Binding of the protein to a ligand induces receptor dimerization and tyrosine autophosphorylation and leads to cell proliferation. This gene is involved in several pathways related to signaling, some type of cancer,  to the cell proliferation, migration and adhesion and to the axon guidance. It is expressed in pediatric brain tumors (\cite{Patereli:2010p17388}).
NTRK2  is member of the neurotrophic tyrosine receptor kinase (NTRK) family. This kinase is a membrane-bound receptor that  upon neurotrophin binding phosphorylates itself and members of the MAPK pathway. Signalling through this kinase leads to cell differentiation. Mutations in this gene have been associated with obesity and mood disorders. SNPs in this gene is associated with AD (\cite{Cozza:2008p17387}).

%% geni AD late
The genes associated to late stage AD are listed in Table \ref{tab:ADsummary} and \ref{tab:alz2late}. 
%The most frequently repeated genes is SNCA. 
Even if SNCA is a known hallmark for PD, it also known to be expressed in late-onset familial AD (\cite{Tsuang:2006p17389}).
Other relevant genes are: SPEN, EIF2AK1, CAT, HBD, ATXN1, XK.
The first gene  a hormone inducible transcriptional repressor. Repression of transcription by this gene product can occur through interactions with other repressors  by the recruitment of proteins involved in histone deacetylation  or through sequestration of transcriptional activators. 
SPEN is involved in the Notch signaling pathway that is important for cell-dell communication since it involves gene regulation mechanisms that control multiple cell differentiation processes (\textit{i.e. neuronal function and development, stabilization of arterial endothelial fate and angiogenesis, cardiac valve homeostasis}) during embryonic and adult life.
EIF2AK1 acts at the level of translation initiation to downregulate protein synthesis in response to stress, therefore it seems to have a protective role diminishing the overproduction of proteins such as SNCA or beta amyloid.
CAT encodes for catalase a key antioxidant enzyme in the bodies defense against oxidative stress, therefore it act against the oxidative stress present in the brain of AD patients. This gene together with EIF2AK1 seems to fight against the disease.
HBD like, HBB commented in subsection \ref{ssec:PD}, could display the same role (\cite{Atamna:2006p17383}).
ATXN1 i s involved in the autosomal dominant cerebellar ataxias (ADCA), an heterogeneous 
group of neurodegenerative disorders characterized by progressive degeneration of the cerebellum brain stem and spinal cord. 
Therefore, because of specific characteristics of these diseases (like the affected brain areas  and the characteristics of the movement disorders),  it might as well play a role in AD.
Finally, mutations of XK have been associated with McLeod syndrome  an X-linked recessive disorder characterized by abnormalities in the neuromuscular and hematopoietic systems.

\section{Conclusion}

The theory of complex networks has recently proven to be a helpful tool for a systematic and structural knowledge of the cell mechanisms. 
Here we propose to enhance its capabilities by coupling it with a machine learning driven approach aimed at moving from a global to a local interaction scales, that is, focussing on pathways which are most likely to change, for instance within particular pathological stages.
Such strategy is also better tailored to deal with situations where small sample size may affect the reliability of the network inference on a global scale.
The method, demonstrated on three disease datasets of environmental pollution, PD and AD, was able to detect biologically meaningful differential pathways.

\clearpage
\begin{small}
\section{Appendix}

\subsection{Experimental setup for the examples}
\label{sec:setup}

The presented analysis pipeline is independent from the single
algorithms chosen for each step of the workflow. Here we give some
details about the methods used for the experiments described in
Section~\ref{sec:discussion}.

\subsubsection{Spectral Regression Discriminant Analysis (SRDA).}
SRDA belongs to the Discriminant Analysis algorithms family (\cite{cai08srda}). 
Its peculiarity is to exploit the regression framework for improving the computational efficiency. 
Spectral graph analysis is used for solving only a set of regularized least squares 
problems avoiding the eigenvector computation.
A score is assigned to each feature and can be interpreted as a feature weight, 
allowing directly feature ranking and selection. 
The regularization value $\alpha$ is the only parameter needed to be tuned.
The method is implemented in Python and it is available within the \texttt{mlpy} library \footnote{\href{http://mlpy.fbk.eu/}{http://mlpy.fbk.eu/}}.

\subsubsection{The $\ell_1\ell_2$ feature selection framework ($\ell_1\ell_2$$_{FS}$).}
$\ell_1\ell_2$$_{FS}$ with double optimization is a feature selection method that can be tuned to give a minimal set of  discriminative  genes or larger sets including correlated genes (\cite{Zou:2005,  DeMol:2009b}). 
The  objective function is a linear model $f(x)=\beta x$,  whose sign gives the classification
rule that can be used to associate a new sample to one of the two classes. 
The sparse weight vector $\beta$ is found by minimizing the $\ell_1\ell_2$ functional: 
$||Y-\beta X||_2^2 + \tau||\beta||_1 +\mu||\beta||_2^2$
where the least square error is penalized with the $\ell_1$ and $\ell_2$ norm of the coefficient vector $\beta$. 
%The least square term ensures fitting of the data whereas adding the two penalties allows to avoid over-fitting. 
The training for selection and classification requires a careful choice of the regularization parameters for both $\ell_1\ell_2$ and RLS. % denoted with $\tau$* and $\lambda$*, respectively.
Indeed, model selection and statistical significance assessment is performed within two nested $K$-cross validation loops as in \cite{Fardin:2009}. 
%The framework provides a set of $K$ lists of selected variables, therefore it  is necessary to choose an appropriate 
%criterion (\cite{Jurman:2008}) in order to assess a common list of relevant
%variables. We based ours on the absolute frequency, {\em i.e.} we decided to promote as
%relevant variables the most stable probesets across the lists.  
The framework is implemented in Python and uses the {\tt L1L2Py} library\footnote{\href{http://slipguru.disi.unige.it/Research/L1L2Py}{http://slipguru.disi.unige.it/Research/L1L2Py}}.

\subsubsection{Functional Characterization.}
The Gene Set Enrichment Analysis (GSEA) was performed by using WebGestalt, an online toolkit\footnote{\href{http://bioinfo.vanderbilt.edu/webgestalt/}{http://bioinfo.vanderbilt.edu/webgestalt/} (\cite{Zhang:2005p4947})}.
This web-service takes as input a list of relevant genes/probesets and performs a  GSEA analysis (\cite{subramanian05gene}) in Kyoto Encyclopedia of Genes and Genomes (KEGG, \cite{Kanehisa:2000}) and Gene Ontology (GO, \cite{Ashburner:2000}),  identifying 
the most relevant pathways and ontologies in the signatures. 
Both for KEGG and GO we selected  the WebGestalt human genome as reference set, $0.05$ as level of
significance, $3$ as the minimum number of genes and the default Hypergeometric test as statistical method.
%Medline \cite{MEDLINE} is used to retrieve the available domain knowledge on the genes. 

\subsubsection{Weighted Gene Co-Expression Networks (WGCN).}
WGCN networks  are based on the idea of using (a function of) the absolute correlation between the expression of a couple of genes across the samples to define a link between them. 
Soft thresholding techniques are then employed to obtain a binary adjacency matrix, where a suitable biologically motivated criterion (such as the scale-free topology, or some other prior knowledge) can be adopted (\cite{zhang05general,zhao10weighted}).
Due to the very small sample size, scale-freeness can not be considered as a reliable criterion 
for threshold selection so we adopted a different heuristics: for both networks in the two classes the
selected threshold is the one maximising the average Ipsen-Mikhailov distance on the selected pathways.

\subsubsection{Algorithm for the Reconstruction of Accurate Cellular Networks (ARACNE).}
ARACNE is a recent method for
inferring networks from the transcription level (\cite{margolin06aracne}) to the metabolic level (\cite{nemenman07reconstruction}).
Beside it was originally designed for handling the complexity of regulatory networks in mammalian cells, 
it is able to address a wider range of network deconvolution problems. 
This information-theoretic algorithm removes the vast majority of indirect candidate interactions inferred by co-expression methods by using the data processing inequality property (\cite{cover91elements}).
In this work we use the MiNET (Mutual Information NETworks) Bioconductor package keeping the default value for the data processing inequality tolerance parameter (\cite{meyer08minet}).
The adopted threshold criterion is the same as the one applied for WGCN.

\subsubsection{Ipsen-Mikhailov distance $\epsilon$.}\label{ssec:distance}
Although already fruitfully used even in a biological context (\cite{sharan06modeling}), the problem of quantitatively comparing network (\textit{e.g.} using a metric instead of evaluating network properties) is a widely open issue affecting many scientific disciplines. 
As discussed in (\cite{jurman10introduction}), many classical distances (such as those of the edit family) have a relevant drawback in being local, that is focussing only on the portions of the network interested by the differences in the presence/absence of matching links. 
%On the other hand, more recent 
More recently, other metrics can overcome this problem so to consider the global structure of the compared topologies; among such distances, the spectral ones - based on the list of eigenvalues of the laplacian matrix of the underlying graph - are quite interesting, and, in particular, the Ipsen-Mikhailov (\cite{ipsen02evolutionary}) distance has been proven to be the most robust in a wide range of situations.

The definition of the $\epsilon$ metric follows the dynamical
interpretation of a $N$-nodes network as a $N$-atoms molecules
connected by identical elastic strings, where the pattern of
connections is defined by the adjacency matrix of the corresponding
network (\cite{ipsen02evolutionary}).
The vibrational frequencies $\omega_i$ of the dynamical system are given by the eigenvalues of the Laplacian matrix of the network: $\lambda_i = -\omega^2_i$, with $\lambda_0=\omega_0=0$.
The spectral density for a graph as the sum of Lorentz distributions is defined as 
$\displaystyle{
\rho(\omega)=K\sum_{i=1}^{N-1} \frac{\gamma}{(\omega-\omega_k)2+\gamma2}
}$, 
where $\gamma$ is the common width\footnote{$\gamma$ specifies the half-width at half-maximum (HWHM), equal to half the interquartile range.} and $K$ is the normalization constant solution of $\int_0^\infty \rho(\omega)\textrm{d}\omega =1$.
Then the spectral distance $\epsilon$ between two graphs $G$ and $H$ with densities $\rho_G(\omega)$ and $\rho_H(\omega)$ can then be defined as 
$
\sqrt{\int_0^\infty \left[\rho_G(\omega)-\rho_H(\omega)\right]^2 \textrm{d}\omega}\ .
$
To get a meaningful comparison of the value of $\epsilon$ on pairs of networks with different number of nodes, we define the normalized version 
$\displaystyle{
\hat{\epsilon}(G,H)=\frac{\epsilon(G,H)}{\epsilon(F_n,E_n)}
}$, 
where $E_n$, $F_n$ indicate respectively the empty and the fully connected network on $n$ nodes: they are the two most $\epsilon$-distant networks for each $n$.
The common width $\gamma$ is set to $0.08$ as in the original reference: being a multiplicative factor, it has no impact on comparing different values of the Ipsen-Mikhailov distance.
The network analysis phase is implemented in R through the \textit{igraph} package.

\subsection{Experiments}\label{sec:exps-supp}

\subsubsection{Air Pollution Experiment}\label{ssec:pollution-supp}

Table \ref{tab:poll1} lists the 11 enriched pathways identified during
the analysis of the air pollution dataset and the total number of the
genes belonging to each pathway. The list is ranked by the normalized
Ipsen-Mikhailov distance $\hat{\epsilon}$ (see Section
\ref{ssec:distance}): the top elements of the list are the most
disrupted pathways between the two conditions. The pathways listed in
Table~\ref{tab:Pollsummary} are a subset of those reported in Table~\ref{tab:poll1}.

Most of these pathways concern the
\textit{developmental} processes: this GO class contains ontologies
especially related to the development of skeletal and nervous systems
(GO:0001501 and GO:0007399) that undergo a rapid and constant growth
in children.  Other enriched terms are related to the capacity of an
organism to defend itself (i.e \emph{response to wounding}, GO:0009611
and \emph{inflammatory response}, GO:0006954), to the regulation of
the cell death (i.e. \emph{negative regulation of apoptosis},
GO:0043066), the \emph{multicellular organismal process}, GO:0032501,
the \emph{glycerlolipid metabolic process}, GO:0046486, the response
to external stimuli (i.e. \emph{inflammatory response}, \emph{response
  to wounding}) and to the locomotion (i.e. GO:0040011, GO:0007626).
\begin{table}[ht]
  \caption{Air Pollution Experiment: pathways corresponding to mostly
    discriminant genes $g_1,...,g_k$ ranked by the normalized Ipsen-Mikhailov
    distance $\hat{\epsilon}$. The number of genes belonging to the pathway is also provided.} \label{tab:poll1}
\begin{center}
{\begin{tabular}{ccc}\toprule 
Pathway & $\hat{\epsilon}$ & \multicolumn{1}{c}{\# Genes} \\ 
\midrule
 GO:0043066 & 0.257 &  21 \\
 GO:0001501 & 0.149 &  89 \\ 
 GO:0009611 & 0.123 &  16 \\
 GO:0007399 & 0.093 & 252 \\
 GO:0016787 & 0.078 & 718 \\
 GO:0005516 & 0.076 & 116 \\ 
 GO:0007275 & 0.076 & 453 \\
 GO:0006954 & 0.048 & 180 \\
 GO:0005615 & 0.038 & 417 \\
 GO:0007626 & 0.000 &   5 \\
 GO:0006066 & 0.000 &   8 \\ 
\botrule
\end{tabular}}{}
\end{center}
\end{table}

Table \ref{tab:poll2} provides the subset of Agilent probesets (together
with their corresponding Gene Symbol and GO pathway) belonging to the
signature $g_1, ..., g_k$ and having a non zero value of the
differential node degree $\Delta d$. Since the $\Delta d$ score is
computed as the difference between the weighted degree in the two
classes, the top elements in Table \ref{tab:poll2} are those whose
number of interactions varies most between the two conditions.

%In Table \ref{tab:poll2} are reported the details of the network analysis results shown in Figure 2 of the main paper.

\begin{table}[ht]
\caption{Air Pollution Experiment: 
list of Agilent probesets in the signature with their corresponding
Entrez Gene Symbol ID and GO pathway.
The list is ranked according to the decreasing absolute value of the
differential node degree $\Delta d$.}\label{tab:poll2}
\begin{center}
{\begin{tabular}{lccr}\toprule 
\multicolumn{1}{c}{Agilent ID}& Gene Symbol & Pathway &
\multicolumn{1}{c}{$\Delta d$}\\ 
\midrule
 4701 &NRGN&GO:0007399 & -2.477 \\
 12235 & DUSP15&GO:0016787 & -1.586 \\ 
 8944 & CLC&GO:0016787 & -1.453 \\ 
 3697 & ITGB5&GO:0007275 & -1.390 \\
 4701 & NRGN&GO:0005516 & -1.357 \\
 12537 &PROK2 &GO:0006954 & 1.069 \\ 
 13835 &OLIG1&GO:0007275 & 0.834 \\
 11673 &HOXB8&GO:0007275 & -0.750 \\
 16424 &FKHL18&GO:0007275 & -0.685 \\
 13094 &DHX32&GO:0016787 & -0.575 \\
 8944 &CLC&GO:0007275 & 0.561 \\ 
 14787 &MATN3&GO:0001501 & 0.495 \\
 15797 &CXCL1 &GO:0006954 & 0.467 \\
 15797 & CXCL1&GO:0005615 & 0.338 \\
 11302 & MYH1&GO:0005516 & -0.194 \\
 15797 &CXCL1 &GO:0007399 & 0.131 \\
\botrule
\end{tabular}}{}
\end{center}
\end{table}

\clearpage
\subsubsection{Parkinson's Disease Experiment}
\label{ssec:PD-supp}

Table \ref{tab:park} reports the list of the pathways selected by the
presented approach both for the early and late PD case, ranked by the normalized
distance $\hat{\epsilon}$. Some of these pathways are also reported in Table~\ref{tab:PDsummary}.

\begin{table}[!h]
\caption{PD Experiment: selected pathways for early (left) and late
  (right) stage corresponding to mostly discriminant genes
  $g_1,...,g_k$ ranked by the normalized Ipsen-Mikhailov distance $\hat{\epsilon}$. The
    number of genes belonging to the pathway is also provided. In bold, the common pathways.}\label{tab:park}
\begin{center}
{\begin{tabular}{ccc|ccc}
\toprule
\multicolumn{3}{c|}{PD early} & \multicolumn{3}{c}{PD late}\\
Pathway& $\hat{\epsilon}$ & \multicolumn{1}{c|}{\# Genes} & Pathway & $\hat{\epsilon}$ & \multicolumn{1}{c}{ \# Genes} \\ 
%PD early & PD early & PD early & PD late & PD late & PD late\\
\midrule
 GO:0012501 & 0.49 &   4 &    GO:0019226 & 0.31 &  20 \\  
 GO:0005764 & 0.39 & 257 &    GO:0010033 & 0.20 &  30 \\
 GO:0019901 & 0.38 & 116 &    GO:0007611 & 0.16 &  34 \\
 GO:0005506 & 0.38 & 434 &    GO:0030234 & 0.15 &  20 \\
 GO:0008219 & 0.38 & 110 &    GO:0042493 & 0.15 & 109 \\
 GO:0016323 & 0.37 & 111 &    GO:0032403 & 0.12 &  14 \\
 GO:0006952 & 0.37 & 160 &    GO:0019717 & 0.12 &  79 \\
 GO:0046983 & 0.36 & 153 &    GO:0009725 & 0.11 &  27 \\
 GO:0045087 & 0.36 & 112 &    GO:0030424 & 0.10 &  93 \\
 GO:0046914 & 0.35 &  51 &    GO:0005096 & 0.09 & 252 \\
 GO:0016265 & 0.33 &   6 &    GO:0007267 & 0.09 & 264 \\
 GO:0042802 & 0.33 & 473 &    GO:0050790 & 0.09 &  15 \\
 GO:0042803 & 0.32 & 411 &    GO:0019001 & 0.09 &  34 \\
 GO:0050896 & 0.31 & 213 &    GO:0017111 & 0.09 & 157 \\
 GO:0006955 & 0.31 & 778 &    GO:0007585 & 0.09 &  47 \\
 GO:0006915 & 0.31 & 687 &    GO:0005516 & 0.09 & 215 \\
 GO:0042981 & 0.30 & 206 &    GO:0005626 & 0.09 &  41 \\
 GO:0030218 & 0.29 &  33 &    GO:0045202 & 0.08 & 278 \\
 GO:0006950 & 0.28 & 253 &    GO:0007610 & 0.08 &  40 \\
 GO:0020037 & 0.26 & 176 &    GO:0005624 & 0.08 & 616 \\
 GO:0005938 & 0.26 &  50 &    GO:0043087 & 0.08 &  22 \\
 GO:0005856 & 0.24 & 816 &    {\bf GO:0003779} & 0.08 & 423 \\
 GO:0016567 & 0.23 & 103 &    GO:0008047 & 0.07 &  60 \\
 {\bf GO:0003779} & 0.23 & 431 &    GO:0042995 & 0.07 & 231 \\
 GO:0042592 & 0.22 &   9 &    GO:0006928 & 0.07 & 166 \\
 GO:0051607 & 0.21 &  26 &    GO:0003924 & 0.07 & 294 \\
 GO:0016564 & 0.18 & 229 &    GO:0007568 & 0.06 &  35 \\
 GO:0005200 & 0.16 & 127 &    GO:0043234 & 0.06 & 233 \\
 GO:0030097 & 0.15 &  76 &    GO:0007268 & 0.06 & 201 \\
 GO:0009615 & 0.14 & 111 &    GO:0030030 & 0.05 &  27 \\
 GO:0008092 & 0.12 &  77 &    GO:0005525 & 0.05 & 450 \\
 GO:0030099 & 0.07 &  19 &    GO:0006412 & 0.05 & 466 \\
 GO:0019900 & 0.04 &  32 &    GO:0043005 & 0.05 &  51 \\
 GO:0034101 & 0.00 &   8 &    GO:0006836 & 0.05 &  42 \\
 GO:0051707 & 0.00 &   5 &    GO:0043025 & 0.04 &  82 \\
\multicolumn{3}{c|}{} &GO:0042221 & 0.00 &  16 \\
\multicolumn{3}{c|}{}& GO:0009266 & 0.00 &   6 \\
\multicolumn{3}{c|}{}& GO:0014070 & 0.00 &  13 \\
\multicolumn{3}{c|}{}& GO:0046578 & 0.00 &   8 \\
\multicolumn{3}{c|}{}& GO:0050804 & 0.00 &  11 \\
\multicolumn{3}{c|}{}& GO:0017076 & 0.00 &   7 \\
\botrule
\end{tabular}}{}
\end{center}
\end{table}

The only common pathway between early and late stage PD is \textit{actin binding} (GO:0003779), as commented in Section~\ref{ssec:PD}. 
% 
%have a very general meaning
%(e.g. \textit{intracellular part, cytoplasm, negative regulation of
%  biological process} GO:0044424, GO:0005737,GO:0048519).
The specific ones for the early stage PD concern the immune system
(i.e. GO:0045087, GO:0006955), the response to stimulus
(i.e. \textit{stress or other organism like virus}, GO:0006950,
GO:0009615), the regulation of metabolic processes, the biological
quality and cell death.  The specific pathways for late stage are
related to the nervous system (e.g. \textit{neurotransmitter
  transport, transmission of nerve impulse, learning or memory},
GO:0006836, GO:0019226) and to response to stimuli
(e.g. \textit{behavior, temperature, organic substances, drugs or
  endogenous stimuli}).

Figure \ref{fig:PDGO} visualizes the enriched pathways in the Molecular Function 
and Biological Process domains. Despite only one pathway was found as 
common between  early and late AD, it is easy to note that the majority of 
selected pathways belong to common GO classes.

%
%
%
%
%Figure \ref{fig:PDGO} represents XXXXXXXXXXX.
%
\begin{figure}[!h]
\begin{center}
\begin{tabular}{cc}
%\centerline{
\includegraphics[width=0.45\textwidth]{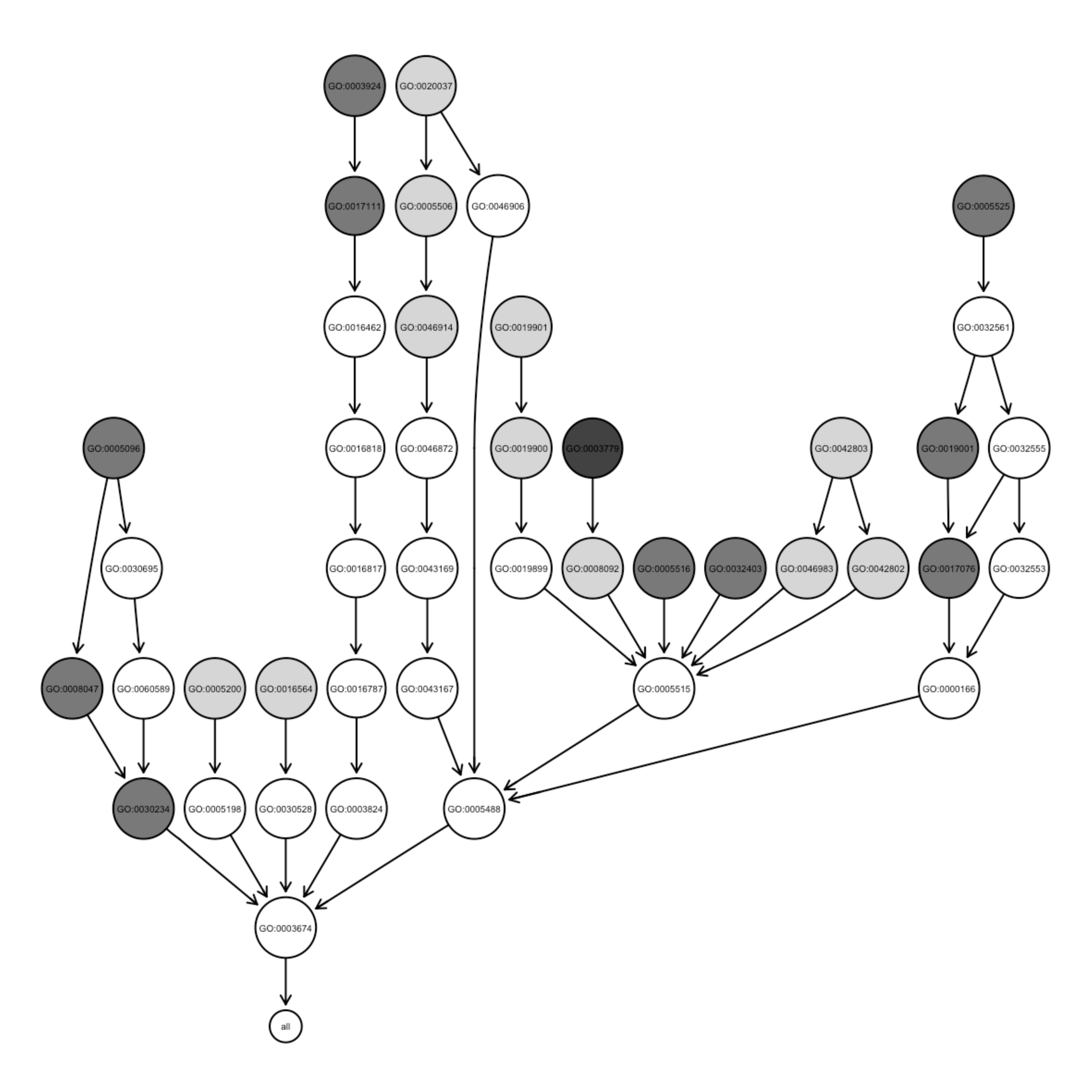} &
\includegraphics[width=0.45\textwidth]{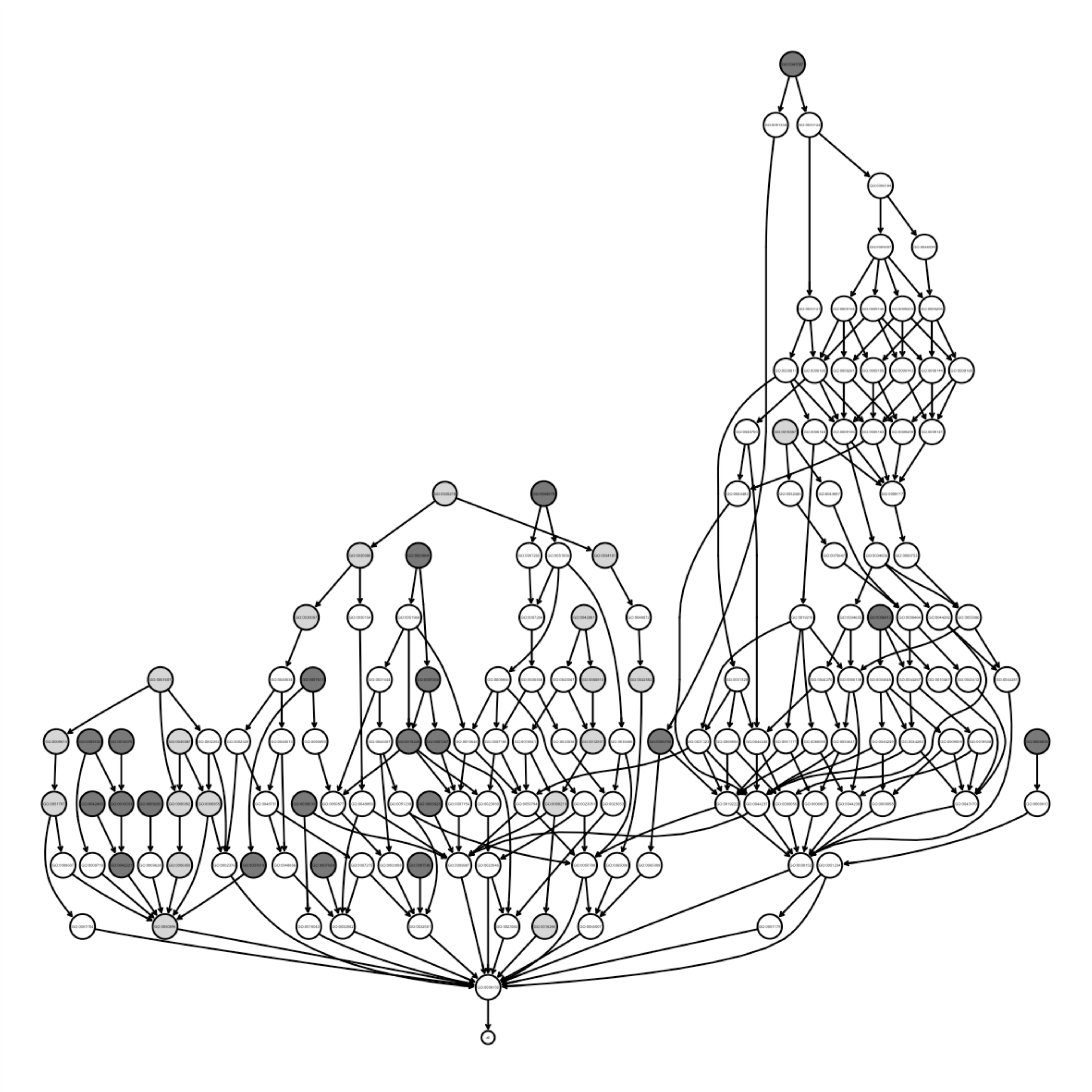} \\
%\includegraphics[width=0.3\textwidth]{./fig/plotParkinsonCC.pdf} \\
%}
(a) MF & (b)  BP \\%& CC (c)\\
\end{tabular}
\caption{GO subgraphs for Parkinson's early  and late stage (Molecular Function and Biological Processes domains). Selected nodes are represented in light gray, gray and dark gray 
for late, early and common nodes.}\label{fig:PDGO}
\end{center}
\end{figure}

Tables \ref{tab:park2early} and \ref{tab:park2late} list respectively
the subset of elements of the early PD signature and late PD signature
having non zero differential node degree $\Delta d$. We recall here
that the top elements in the two tables are those whose number of
interactions varies most between the two case/control conditions.
\begin{table}[!h]
  \caption{PD Experiment (early): list of Affymetrix probesets in
    the early stage signature with their corresponding Entrez Gene
    Symbol ID and GO pathway. The list is ranked according to the decreasing absolute value of the differential node degree $\Delta d$.}\label{tab:park2early}
\begin{center}
{\begin{tabular}{lccr}
\toprule
\multicolumn{1}{c}{Affy Probeset ID} & Gene Symbol & Pathway & \multicolumn{1}{c}{$\Delta d$}\\
\midrule
200931\_s\_at & VCL         & GO:0005856 & -2.124 \\ 
200931\_s\_at & VCL         & GO:0003779 & -2.107 \\ 
213067\_at    & MYH10       & GO:0003779 & 1.879 \\ 
202887\_s\_at & DDIT4       & GO:0006915 & -1.872 \\ 
201841\_s\_at & HSPB1       & GO:0042802 & -1.691 \\ 
204439\_at    & IFI44L      & GO:0006955 & -1.585 \\ 
201841\_s\_at & HSPB1       & GO:0005856 & -1.532 \\ 
209480\_at    & HLA-DQB1    & GO:0006955 & -1.340 \\ 
209116\_x\_at & HBB         & GO:0005506 & -1.008 \\ 
203998\_s\_at & SYT1        & GO:0042802 & 0.862 \\ 
36711\_at     & MAFF        & GO:0006950 & -0.807 \\ 
209116\_x\_at &HBB          & GO:0020037 & -0.599 \\ 
36711\_at     & MAFF        & GO:0046983 & 0.567 \\
214414\_x\_at &HBA1/HBA2    & GO:0020037 & -0.376 \\ 
205033\_s\_at &DEFA1/DEFA3  & GO:0009615 & -0.360 \\ 
217232\_x\_at &HBB          & GO:0005506 & -0.239 \\ 
205033\_s\_at &DEFA1/DEFA3  & GO:0006955 & -0.228 \\ 
213067\_at    & MYH10       & GO:0005938 & 0.183 \\
214414\_x\_at &HBA1/HBA2    & GO:0005506 & -0.182 \\ 
201841\_s\_at &HSPB1        & GO:0006950 & -0.154 \\ 
217232\_x\_at &HBB          & GO:0020037 & -0.088 \\ 
218197\_s\_at &OXR1         & GO:0006950 & -0.059 \\
205033\_s\_at & DEFA1/DEFA3 & GO:0006952 & 0.027 \\ 
\botrule
\end{tabular}}{}
\end{center}
\end{table}

In Table \ref{tab:park2early}, we note that the most disrupted genes for early PD 
(IFI44L, HSPB1, MAFF, DEFA1/DEFA3, OXR1)
belong to pathways related to  {\em response to stress} and {\em to virus}. 
Moreover, several genes (HLA-DQB1, HBB, HBA1/HBA2, DEFA1/DEFA3) 
are related to the following pathways: {\em heme binding}, {\em iron ion binding} and {\em immune response}.

In Table \ref{tab:park2late} the majority of disrupted genes for late PD
(RGS4, CDC42, RABGAPIL) 
occur in pathways that are related to GTP, a purine nucleotide that
can function either as source of energy for protein synthesis and in
the signal transduction particularly with G-proteins.  Other genes 
in the list (MYH10, RGS4, PHACTR1, SYT1, VCL) 
are related to the acting and calmodulin binding, to the
synaptic transmission, the neurotransmitter transport, the cell-cell
signaling, the translation and the cellular component movement.  
The majority of the disrupted pathways are located in several components
of the neurons.

\begin{table}[!h]
\caption{PD Experiment (late): list of Affymetrix probesets in
  the late stage signature with their corresponding Entrez Gene Symbol and GO pathway.
The list is ranked according to the decreasing absolute value of the differential node degree $\Delta d$.}\label{tab:park2late}
\begin{center}
{\begin{tabular}{lccr}
\toprule
\multicolumn{1}{c}{Affy Probeset ID} & Entrez Gene Symbol & Pathway & \multicolumn{1}{c}{$\Delta d$}\\
\midrule
213638\_at & PHACTR1 & GO:0045202 & -3.255 \\ 
213067\_at & MYH10 & GO:0003779 & -3.252 \\ 
213067\_at & MYH10 & GO:0005516 & -2.597 \\ 
204337\_at & RGS4 & GO:0005096 & -2.194 \\ 
213067\_at & MYH10 & GO:0043025 & -2.107 \\ 
213067\_at & MYH10 & GO:0043005 & -1.696 \\ 
214230\_at & CDC42 & GO:0003924 & 1.677 \\ 
213638\_at & PHACTR1 & GO:0003779 & -1.587 \\ 
213067\_at & MYH10 & GO:0030424 & -1.170 \\ 
205857\_at & SLC18A2 & GO:0006836 & -1.094 \\ 
206552\_s\_at & TAC1 & GO:0007268 & -0.834 \\ 
206552\_s\_at & TAC1 & GO:0007267 & 0.809 \\ 
205110\_s\_at & FGF13 & GO:0007267 & -0.804 \\ 
203998\_s\_at & SYT1 & GO:0005516 & -0.787 \\ 
208319\_s\_at & RBM3 & GO:0006412 & -0.759 \\ 
201909\_at & RPS4Y & GO:0006412 & -0.688 \\ 
200931\_s\_at & VCL & GO:0043234 & -0.655 \\ 
204337\_at & RGS4 & GO:0005516 & -0.602 \\ 
205105\_at & MAN2A1 & GO:0007585 & -0.502 \\ 
205857\_at & SLC18A2 & GO:0005624 & -0.428 \\ 
201841\_s\_at & HSPB1 & GO:0006928 & 0.424 \\ 
214230\_at & CDC42 & GO:0042995 & -0.379 \\ 
214230\_at & CDC42 & GO:0005525 & 0.370 \\ 
203998\_s\_at & SYT1 & GO:0043005 & -0.357 \\ 
 203998\_s\_at & SYT1 & GO:0045202 & -0.339 \\ 
 200931\_s\_at & VCL & GO:0003779 & -0.311 \\ 
 200931\_s\_at & VCL & GO:0006928 & -0.308 \\ 
 206836\_at & SLC6A3 & GO:0006836 & -0.238 \\ 
 215342\_s\_at & RABGAP1L & GO:0005096 & -0.211 \\ 
 211727\_s\_at & COX11 & GO:0007585 & 0.188 \\ 
 203998\_s\_at & SYT1 & GO:0007268 & -0.159 \\ 
\botrule
\end{tabular}}{}
\end{center}
\end{table}

%The main findings are summarized in Table 2 of the main paper.
% In Tables \ref{tab:park2early} and \ref{tab:park2late} are reported
% the details of the network analysis results shown in Figure 3 of the
% main paper only for the early stage of PD.

\clearpage
\subsubsection{Alzheimer's Disease Experiment}
\label{ssec:AD-supp}

Table \ref{tab:alz} reports the most discriminant pathways for the two
AD stages as selected by the presented pipeline, ranked by decreasing normalized
$\hat{\epsilon}$ distance. 
%Some of these pathways are also reported in Table 3
%of the main paper. 
Table~\ref{tab:ADsummary} summarizes the main results here detailed in
Table \ref{tab:alz},  \ref{tab:alz2early} and \ref{tab:alz2late}. 
\begin{table}[!h]
\caption{AD Experiment: selected pathways for early (left) and late
  (right) stage corresponding to mostly discriminant genes
  $g_1,...,g_k$ ranked by the  normalized Ipsen-Mikhailov distance $\hat{\epsilon}$. The
    number of genes belonging to the pathway is also provided. In bold, the common pathways.}\label{tab:alz}
\begin{center}
{\begin{tabular}{ccc|ccc}
\toprule
\multicolumn{3}{c|}{AD early} & \multicolumn{3}{c}{AD late}\\
Pathway& $\hat{\epsilon}$ & \multicolumn{1}{c|}{\# Genes} & Pathway & $\hat{\epsilon}$ & \multicolumn{1}{c}{ \# Genes} \\ 
%PD early & PD early & PD early & PD late & PD late & PD late\\
\midrule
  GO:0048514 & 0.22 &  22 &GO:0040012 & 0.36 &   9 \\ 
  {\bf GO:0042598} & 0.21 &  16 &{\bf GO:0042598} & 0.23 &  16 \\ 
  GO:0016881 & 0.19 & 109 &{\bf GO:0019226} & 0.12 &  27 \\ 
  GO:0019787 & 0.16 & 116 &GO:0030334 & 0.10 &  93 \\ 
  GO:0019725 & 0.16 &  14 &GO:0045892 & 0.09 & 218 \\ 
  GO:0051246 & 0.15 & 121 &GO:0009968 & 0.06 & 107 \\ 
  GO:0001508 & 0.14 &  31 &GO:0042493 & 0.06 & 160 \\ 
  GO:0006631 & 0.14 & 171 &GO:0050877 & 0.06 &  31 \\ 
  GO:0030234 & 0.13 &  29 &GO:0042127 & 0.05 & 140 \\ 
  GO:0016874 & 0.12 & 735 &GO:0009725 & 0.05 &  47 \\ 
  GO:0004842 & 0.11 & 368 &GO:0042277 & 0.05 &  63 \\ 
  GO:0007417 & 0.10 & 199 &GO:0015630 & 0.05 &  99 \\ 
  GO:0012505 & 0.10 & 216 &GO:0008283 & 0.04 & 785 \\ 
  GO:0050880 & 0.09 &  26 &GO:0005819 & 0.04 & 142 \\ 
  GO:0048471 & 0.08 & 263 &GO:0008217 & 0.03 & 106 \\ 
  GO:0005792 & 0.08 & 409 &GO:0005626 & 0.03 &  68 \\ 
  GO:0005768 & 0.08 & 490 &GO:0000165 & 0.03 &  94 \\ 
  GO:0004857 & 0.08 &  57 &GO:0005215 & 0.03 & 685 \\ 
  GO:0031982 & 0.07 &  34 &GO:0007268 & 0.03 & 377 \\ 
  GO:0016567 & 0.07 & 206 &GO:0007601 & 0.03 & 402 \\ 
  GO:0008217 & 0.07 & 105 &GO:0008289 & 0.03 & 285 \\ 
  GO:0001666 & 0.07 & 225 &GO:0007610 & 0.03 &  84 \\ 
  GO:0030141 & 0.06 &  69 &GO:0008284 & 0.02 & 507 \\ 
  GO:0050877 & 0.06 &  31 &GO:0001503 & 0.02 & 171 \\ 
  GO:0042552 & 0.05 &  36 &GO:0007243 & 0.02 & 220 \\ 
  GO:0001568 & 0.05 &  79 &GO:0008285 & 0.02 & 578 \\ 
  GO:0048511 & 0.04 &  49 &{\bf GO:0008015} & 0.02 & 103 \\ 
  GO:0016023 & 0.04 & 108 &GO:0016564 & 0.02 & 380 \\ 
  GO:0007399 & 0.04 & 806 &GO:0020037 & 0.02 & 265 \\ 
  {\bf GO:0008015} & 0.04 & 103 &GO:0051270 & 0.00 &   9 \\ 
  GO:0042391 & 0.04 &  67 &GO:0010033 & 0.00 &  44 \\ 
  GO:0031410 & 0.03 & 482 &GO:0050890 & 0.00 &  31 \\ 
  GO:0046982 & 0.03 & 364 &GO:0050953 & 0.00 &  24 \\ 
  GO:0006633 & 0.02 & 109 &{\bf GO:0000267} & 0.00 &   5 \\ 
  GO:0045121 & 0.02 & 136 &&\\ 
  GO:0004866 & 0.02 & 194 &&\\ 
  GO:0008366 & 0.00 &  22 &&\\ 
  GO:0019228 & 0.00 &  19 &&\\ 
  GO:0006873 & 0.00 &  10 &&\\ 
  GO:0042592 & 0.00 &  25 &&\\ 
  GO:0001974 & 0.00 &  28 &&\\ 
  {\bf GO:0019226} & 0.00 &  27 &&\\ 
  GO:0001944 & 0.00 &   4 &&\\ 
  GO:0048771 & 0.00 &  12 &&\\ 
  GO:0048856 & 0.00 &  20 &&\\ 
  GO:0019838 & 0.00 &  85 &&\\ 
  GO:0017076 & 0.00 &  11 &&\\ 
  GO:0030414 & 0.00 &  42 &&\\ 
  GO:0001882 & 0.00 &   8 &&\\ 
  {\bf GO:0000267} & 0.00 &   4 &&\\ 
  GO:0031090 & 0.00 &   6 &&\\ 
\botrule
\end{tabular}}{}
\end{center}
\end{table}
%
%For each signature, the  enrichment analysis identified relevant enriched nodes 
%either specific or common between early and late AD. 
The common pathways are: 
GO:0019226 \textit{i.e. transmission of nerve impulse}, GO:0008015 \textit{i.e. blood circulation}, 
GO:0000267 \textit{i.e. cell fraction} and GO:0042598 \textit{i.e. vesicular fraction}. 
The relevance of blood circulatory system in AD has already been highlighted 
in  \cite{Brown:2011p17396} and references therein.

Figure \ref{fig:ADGO} visualizes the enriched pathways in the Molecular Function 
and Biological Process domains. Despite only 4 pathways were found as 
common between  early and late AD, it is easy to note that the majority of 
selected pathways belong to common GO classes.  

\begin{figure}[!h]
\begin{center}
\begin{tabular}{cc}
%\centerline{
\includegraphics[width=0.45\textwidth]{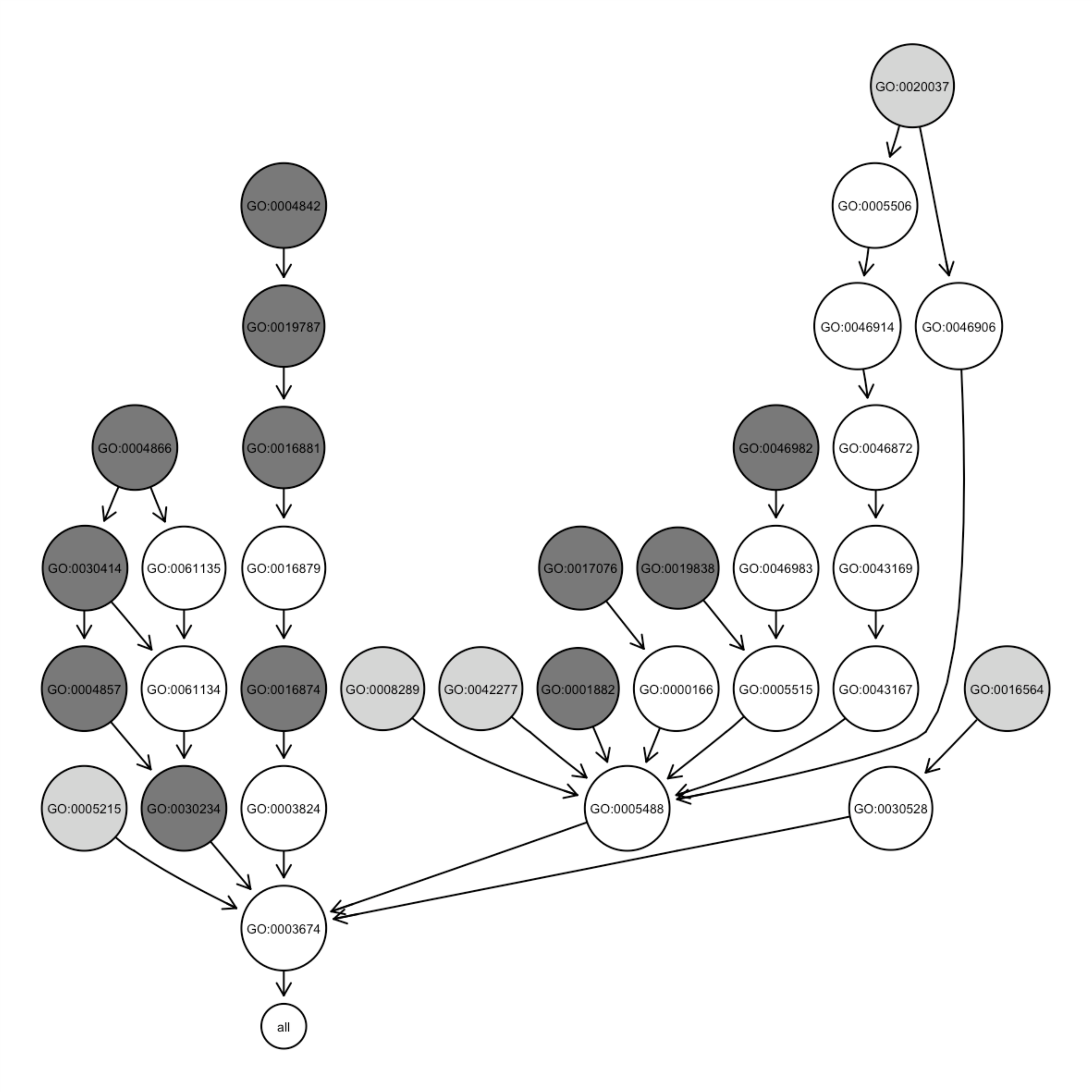} &
\includegraphics[width=0.45\textwidth]{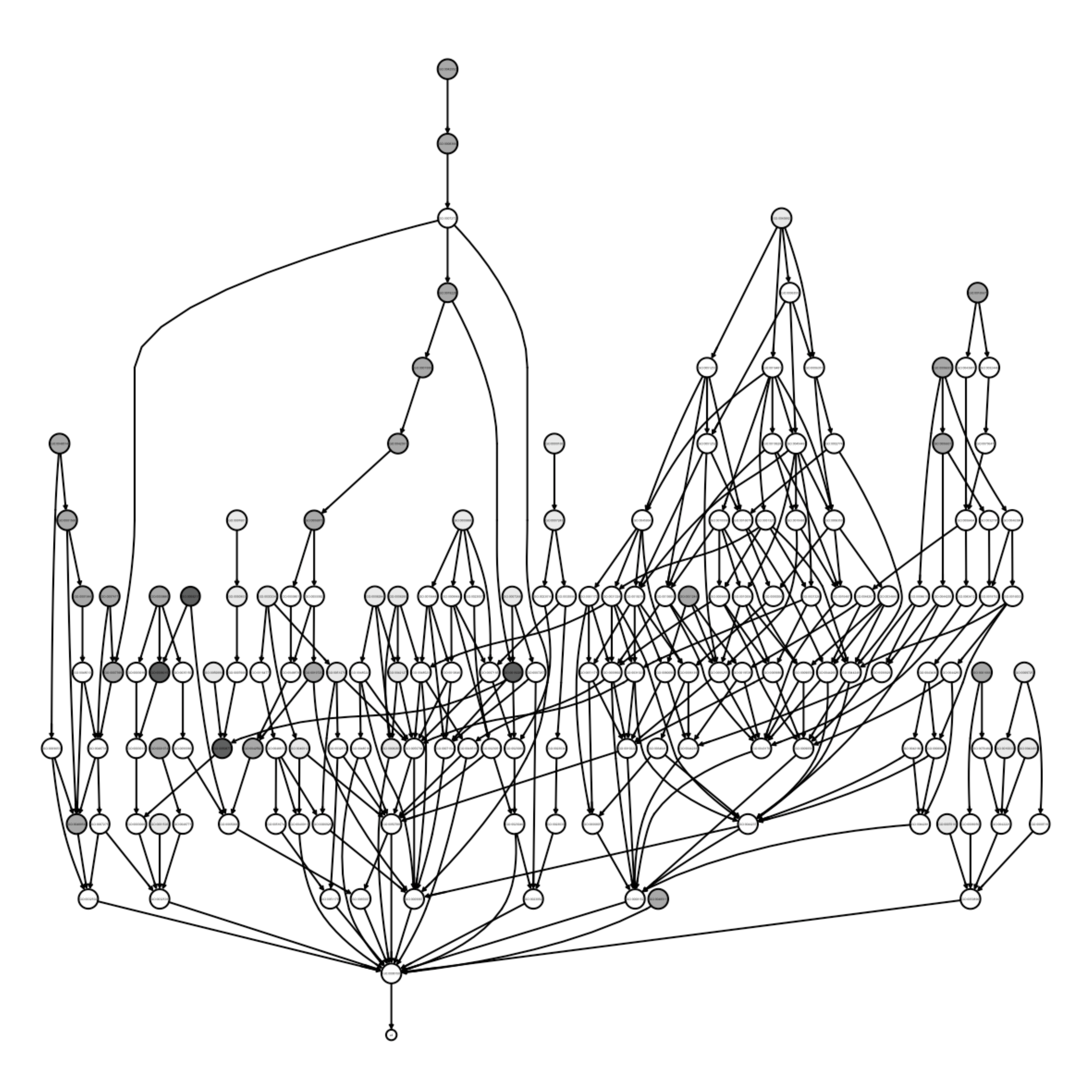} \\
%\includegraphics[width=0.3\textwidth]{./fig/plotParkinsonCC.pdf} \\
%}
(a) MF & (b)  BP \\%& CC (c)\\
\end{tabular}
\caption{GO subgraphs for Alzheimer's early and late stage (Molecular Function and Biological Processes domains). Selected nodes are represented in light gray, gray and dark gray 
for late, early and common nodes.}\label{fig:ADGO}
\end{center}
\end{figure}

%% non metterei contenuto in queste XX sotto riportate perche' sa
%
%The specific ones for the early stage AD XXXXXX.
%
%The specific ones for the late stage AD XXXXXX.
Tables \ref{tab:alz2early} and \ref{tab:alz2late} provide  details of the network analysis results on early and late stage AD, respectively. 
The elements of the two signatures having non zero $\Delta d$ are listed
for decreasing absolute value of the differential node degree score,
thus giving top positions to genes that change most the interaction
network between the case/control condition.

\begin{table}[!h]
\caption{AD Experiment (early): list of Affymetrix probesets in
  the early stage signature with their corresponding Entrez Gene Symbol and GO pathway.
The list is ranked according to the decreasing absolute value of the differential node degree $\Delta d$.}\label{tab:alz2early}
\begin{center}
{\begin{tabular}{llcr}
\toprule
\multicolumn{1}{c}{Affy Probeset ID} &Gene Symbol&Pathway& \multicolumn{1}{c}{$\Delta d$}\\
\midrule
  209116\_x\_at &HBB & GO:0050880 & 1.670\\ 
  209116\_x\_at & HBB& GO:0008217 & 1.445\\ 
  211748\_x\_at &PTGDS& GO:0006633 & 1.273\\ 
  240383\_at    &UBE2D3& GO:0016874 & -1.165\\ 
  240383\_at    &UBE2D3& GO:0019787 & -0.703 \\ 
  201061\_s\_at &STOM& GO:0045121 & -0.662 \\ 
  240383\_at    &UBE2D3& GO:0051246 & -0.613 \\ 
  201983\_s\_at &EGFR& GO:0046982 & -0.476 \\ 
  221795\_at    &NTRK2& GO:0007399 & -0.262 \\ 
  212226\_s\_at &PPAP2B& GO:0001568 & 0.259 \\ 
  201983\_s\_at &EGFR& GO:0005768 & -0.256 \\ 
  211696\_x\_at &HBB& GO:0050880 & -0.224 \\ 
  209072\_at    &MBP& GO:0008366 & 0.166 \\ 
  211696\_x\_at &HBB& GO:0008217 & -0.149 \\ 
  212187\_x\_at &PTGDS& GO:0006633 & -0.139 \\ 
  201185\_at    &HTRA1& GO:0019838 & 0.124 \\ 
  240383\_at    &UBE2D3& GO:0004842 & 0.120\\ 
  209072\_at    & MBP& GO:0007417 & 0.113 \\ 
  240383\_at    & UBE2D3& GO:0016567 & -0.047 \\ 
\botrule
\end{tabular}}{}
\end{center}
\end{table}

%Figure \ref{fig:ADGO} XXXXXXX
%
%\begin{figure*}[!h]
%\centerline{\includegraphics[width=\textwidth]{./fig/plotADMF.png}}
%\caption{GO subgraph for Parkinson's early  and late stage. yellow = late, green= common, red = early.}\label{fig:ADGO}
%\end{figure*}

Table \ref{tab:alz2early} reports the most disrupted probesets within the early stage AD, ranked according to 
the differential node degree $\Delta d$. 
We note that the most disrupted gene is HBB, within {\em regulation of blood vessel size} and {\em regulation of blood vessels}.  % (\cite{Atamna:2006p17383})

\begin{table}[!h]
\caption{AD Experiment (late): list of Affymetrix probesets in
  the late stage signature with their corresponding Entrez Gene Symbol and GO pathway.
The list is ranked according to the decreasing absolute value of the differential node degree $\Delta d$.}\label{tab:alz2late}
\begin{center}
{\begin{tabular}{llcr}
\toprule
\multicolumn{1}{c}{Affy Probeset ID} &Gene Symbol&Pathway& \multicolumn{1}{c}{$\Delta d$}\\
\midrule
  201996\_s\_at &SPEN & GO:0016564 & 1.590\\ 
  211546\_x\_at &SNCA& GO:0040012 & 1.410\\ 
  211546\_x\_at &SNCA& GO:0042493 & 1.310\\ 
  201996\_s\_at &SPEN& GO:0045892 & 1.246\\ 
  217736\_s\_at &EIF2AK1& GO:0020037 & -1.066\\ 
  201005\_at    &CD9& GO:0008285 & 0.725\\ 
  210943\_s\_at &LYST& GO:0015630 & 0.706\\ 
  204466\_s\_at &SNCA& GO:0042493 & 0.461\\ 
  207827\_x\_at &SNCA & GO:0040012 & 0.434\\ 
  206698\_at    &XK& GO:0005215 & 0.433\\ 
  209184\_s\_at &IRS2& GO:0008283 & 0.208\\ 
  212420\_at    &ELF1& GO:0016564 & -0.203\\ 
  207827\_x\_at &SNCA& GO:0042493 & 0.201\\ 
  205592\_at    &SLCA4A1& GO:0005215 & 0.180\\ 
  211922\_s\_at &CAT& GO:0008283 & 0.173\\ 
  211922\_s\_at &CAT& GO:0020037 & -0.094 \\ 
  203231\_s\_at &ATXN1& GO:0016564 & -0.073 \\ 
  217736\_s\_at &EIF2AK1& GO:0008285 & -0.072 \\ 
  204466\_s\_at &SNCA& GO:0040012 & 0.048 \\ 
  206834\_at    &HBD& GO:0008217 & 0.045\\ 
  206834\_at    &HBD& GO:0020037 & 0.019 \\ 
\botrule
\end{tabular}}{}
\end{center}
\end{table}

Table \ref{tab:alz2late} reports the most disrupted genes within the late stage AD, ranked according to 
the differential node degree $\Delta d$. The majority of such genes (SPEN, SNCA, EIF2AK1, ELF1, CAT, ATXN1, HBD) belong to {\em regulation of 
locomotion}, {\em transcription repressor activity}, {\em response to drug} and {\em heme binding}.

\end{small}
\clearpage
\bibliographystyle{natbib}
\bibliography{barla11pathway}

\end{document}